\documentclass[numberedappendix,apj]{emulateapj}
\usepackage{apjfonts} 

\makeindex
\citeindextrue

\shorttitle{The Trails of Superluminal Jet Components in 3C\,111}
\shortauthors{Kadler et al.}
\received{2007 September 26}
\accepted{2007 December 19}
\journalinfo{Astrophysical~journal}
\submitted{}

\begin{document}

\title{The Trails of Superluminal Jet Components in 3C\,111}

\author{M. Kadler\altaffilmark{1,2}
        E. Ros\altaffilmark{2},
        M. Perucho\altaffilmark{2},
        Y. Y. Kovalev\altaffilmark{2,3},
        D. C. Homan\altaffilmark{4},
        I. Agudo\altaffilmark{5,2},
        K. I. Kellermann\altaffilmark{6},
        M. F. Aller\altaffilmark{7},
        H. D. Aller\altaffilmark{7},
        M. L. Lister\altaffilmark{8},
        and
        J. A. Zensus\altaffilmark{2}
        }
\altaffiltext{1}{Astrophysics Science Division, NASA's Goddard Space Flight Center, Greenbelt Road, Greenbelt, MD 20771, USA; \mbox{Matthias.Kadler@nasa.gov}} 
\altaffiltext{2}{Max-Planck-Institut f\"ur Radioastronomie, Auf dem H\"ugel\,69, 53121 Bonn, Germany; ros, perucho, ykovalev, \mbox{azensus@mpifr-bonn.mpg.de}}
\altaffiltext{3}{Astro Space Center of Lebedev Physical Institute, Profsoyuznaya 84/32, 117997 Moscow, Russia}
\altaffiltext{4}{Astronomy Department, Department of Physics and Astronomy, Denison University, Granville, OH 43023, U.S.A.; \mbox{homand@denison.edu}}
\altaffiltext{5}{Instituto de Astrof\'{\i}sica de Andaluc\'{\i}a (CSIC),  Apartado 3004, E-18080 Granada, Spain; \mbox{iagudo@iaa.es}}
\altaffiltext{6}{National Radio Astronomy Observatory, 520 Edgemont Road, Charlottesville, VA 22903, U.S.A; \mbox{kkellerm@nrao.edu}}
\altaffiltext{7}{Astronomy Department, University of Michigan, Ann Arbor, MI 48109-1042, U.S.A.; mfa, \mbox{haller@umich.edu}}
\altaffiltext{8}{Department of Physics, Purdue University, 525 Northwestern Avenue, West Lafayette, IN 47907, U.S.A.; \mbox{mlister@physics.purdue.edu}}

\begin{abstract}
In 1996, a major radio flux-density outburst occured in the broad-line radio galaxy 3C\,111.
It was followed by a particularly bright plasma ejection associated with a superluminal
jet component, which has shaped the parsec-scale structure of 3C\,111 for almost a decade. 
Here, we present results from 18 epochs of Very Long Baseline Array (VLBA) observations
conducted since 1995 as
part of the VLBA 2\,cm Survey and MOJAVE monitoring programs.
This major event allows us to study a variety of processes associated with outbursts of radio-loud
AGN in much greater detail than has been possible in other cases: the primary perturbation gives rise to
the formation of a leading and a following component, which are interpreted as 
a forward and a backward-shock. Both components evolve in characteristically different ways
and allow us to draw conclusions about the work flow of jet-production events; the expansion, acceleration
and recollimation
of the ejected jet plasma in an environment with steep pressure and density gradients are revealed;
trailing components are formed in the wake of the primary perturbation possibly as a result of coupling to 
Kelvin-Helmholtz instability pinching modes from the interaction of the jet with the external medium. The 
interaction of the jet with its ambient medium 
is further described by the linear-polarization signature of jet components traveling
along the jet and passing a region of steep pressure/density gradients. 
\end{abstract}

\keywords{galaxies: individual: 3C111 --
                galaxies: active --
                galaxies: jets --
                galaxies: nuclei }

\section{Introduction}
Direct evidence for the existence of bulk relativistic outflows along the jets in 
blazars and other radio-loud active galactic nuclei (AGN) comes from
Very-Long-Baseline Interferometry (VLBI) observations. 
The first evidence for apparently superluminal 
structural changes was found from changes in the fringe visibility curves
of 3C\,279 and 3C\,273 \citep{Whi71,Coh71}.
Subsequent higher-quality VLBI 
observations \citep[see, e.g., compilation by ][and references therein]{Ver94} have established
the ``core-jet'' type  milliarcsecond-scale structure of compact extragalactic jets:
the core being a bright and unresolved flat-spectrum component at the end of a linear
structure, and the jet being composed out of individual steep-spectrum components or ``knots''.
The knots frequently move away from the core with apparent velocities exceeding the
speed of light.
Monitoring observations of large source samples 
\citep{Ver94,Jor01,Kel04,Pin07}
have provided important statistical tools for probing
relativistic beaming and the intrinsic properties of extragalactic radio jets \citep{Coh07},
their intrinsic brightness temperatures \citep{Hom06}, or their Lorentz factor
distribution \citep{Kel04} and luminosity function \citep{Car07}.

The relativistic-jet model \citep[e.g.,][]{Bla79}
has become the de-facto paradigm in multiwavelength research on blazars and other AGN,
but VLBI observations 
have demonstrated that the basic concept of ballistically-moving isolated jet knots
is clearly oversimplified: 
jet curvature \citep[e.g.,][]{Ver94}, stationary components \citep[e.g.,][]{Jor01}, and 
non-radial and accelerated motions \citep[e.g.,][]{Kel04}, are found to be common
features of relativistic jets. 
Within individual jets, there are often characteristic velocities 
suggesting the presence of an underlying continuous jet flow, but the ``components''
themselves most likely represent patterns moving  at a different speed than the underlying
flow, e.g., as hydrodynamically propagating shocks \citep{Mar85,Hug85}.  

Recent years have brought major improvements in numerical simulations of relativistic jets
\citep[see, e.g.,][for a review]{Gom05}.
It is now possible to simulate three-dimensional 
relativistic jets \citep[e.g.,][]{Alo03} and to compute 
the relativistic processes \citep[e.g.,][]{Gom97} that transfer hydrodynamic
results into observed brightness distributions (e.g., relativistic light abberation and
light travel time delays).
In particular, 
interactions between strong perturbations or shocks with the underlying jet flow and the 
jet-ambient medium can be simulated \citep{Agu01}. 
With these new techniques, it is now possible to compare the generation, propagation and evolution
of emission features in simulated and observed relativistic jets. 

The nearby (z$=$0.049)\footnote{Assuming $H_0=71$\,km\,s$^{-1}$\,Mpc$^{-1}$,
$\Omega_{\rm M} = 0.3$, $\Omega_\Lambda = 0.7$ ($1{\rm \,mas} = 1.0{\rm \,pc}$).} broad-line radio galaxy 3C\,111 (PKS B\,0415+379)
shows a classical FR\,II morphology on kiloparsec-scales
spanning more than 200$^{\prime\prime}$ with a highly collimated jet connecting
the central core and the northeastern lobe in position angle 63$^\circ$ while
no counterjet is observed towards the southwestern lobe \citep{Lin84}. This asymmetry
is usually explained via relativistic boosting of the jet and de-boosting of
the counter-jet.
3C\,111 exhibits the brightest compact radio core at cm/mm wavelengths of all
FR\,II radio galaxies, a blazar-like spectral energy distribution \citep{Sgu05}, and
it was one of the first (and only) radio galaxies in which superluminal motion
was detected \citep{Goe87,Pre88}.
Moreover, the (sub-) parsec
scale jet of 3C 111 is intimately related to its high-energy emission: \citet{Mar06}
reports a disk-jet connection, similar to the well-established one in 3C 120 \citep{Mar02}, in the sense that dips in the X-ray
light curve indicate accretion events which are followed by VLBI jet component ejections.
Recently, R.C.\,Hartman \& M. Kadler (in prep.) showed that
the gamma-ray source 3EG\,J0416+3650 can be decomposed into multiple individual
sources inside the EGRET full-band point-spread function,
revealing a significant signal from the nominal position of 3C\,111
in the higher-resolution, high-energy band above 1\,GeV. 
This association of 3C\,111 with 3EG\,J0416+3650, which had originally been suggested 
by \citet{Har99} and \citet{Sgu05},
makes 3C\,111 one of the very rare radio galaxies detected
at gamma-ray energies and supports the view that this 
source may be considered a lower-luminosity version of powerful radio-loud quasars. 

Here, we report
the results from ten years of Very-Long-Baseline Interferometry (VLBI) observations of
3C\,111 as part of 
the  VLBA 2\,cm Survey\footnote{\tt http://www.cv.nrao.edu/2cmsurvey/} \citep{Kel98,Zen02,Kel04,Kov05} and its follow-up program 
MOJAVE\footnote{\tt http://www.physics.purdue.edu/MOJAVE} \citep{Lis05,Hom06b}.
We investigate the parsec-scale source
structure during a major flux-density outburst and during its aftermath.
We find that this outburst was associated with the formation of an exceptionally bright  feature
in the jet of 3C\,111. 
A variety of processes (beyond the predictions of simple ballistic motion models) are observed and
discussed in view of modern relativistic-jet simulations.
In Sect.~\ref{sect:obs}, our observations
and the data reduction are described. A detailed report of the observational results is given in
Sect.~\ref{sect:results}. In Sect.~\ref{sect:discussion}, we discuss the various
processes observed in the jet of 3C\,111 as a result of
the outburst and during the propagation of the new jet feature along the jet. In Sect.~\ref{sect:conclusions}, 
we put these results into the context of future simulations and observations
with the goal of understanding the production mechanisms of AGN jets.

\section{Observations and data analysis}
\label{sect:obs}
3C\,111 has been monitored as part of the VLBA 2\,cm Survey program since
April 1995.
The observational details are given by
\citet{Kel98}.
Following the methods described  there,
the data from 17 epochs of VLBA observations of 3C\,111
between 1995 and 2005 (see Table~\ref{tab:3c111_journal})
were phase and amplitude self calibrated
and the brightness distribution was determined
via hybrid mapping. 
An additional epoch from June 2000 was made available to us by 
G. Taylor.
The polarization calibration was performed as described in \citet{Lis05}.
Two-dimensional Gaussian components  were fitted in the $(u,v)$-domain
to the fully calibrated visibility data of each epoch
using the program {\sc difmap} \citep{She97}.
The parameters of each model fit at the various epochs are
given in Table~\ref{tab:3c111-modelfits}.
The models were aligned by assuming the
westernmost component (namely, the ``core'') to
be stationary so that the position of jet components can be measured relative
to it.
Because of the coupling of the flux densities of nearby model components,
the uncertainties in the component flux densities are larger than the formal (statistical) errors
unless the given model component is far enough 
from its closest neighbor. Throughout this paper, errors of
15\,\% are assumed for the flux densities of individual model-fit
components. 
In most cases, this should be considered a conservative estimate that accounts for absolute
calibration uncertainties and formal model-fitting uncertainties \citep[see, e.g.,][]{Hom02}. 
Position uncertainties were determined
internally from the deviations of the data from linear motion.

\section{Results}
\label{sect:results}

\begin{figure}[t]
   \centering
   \includegraphics[clip,angle=-90,width=\linewidth]{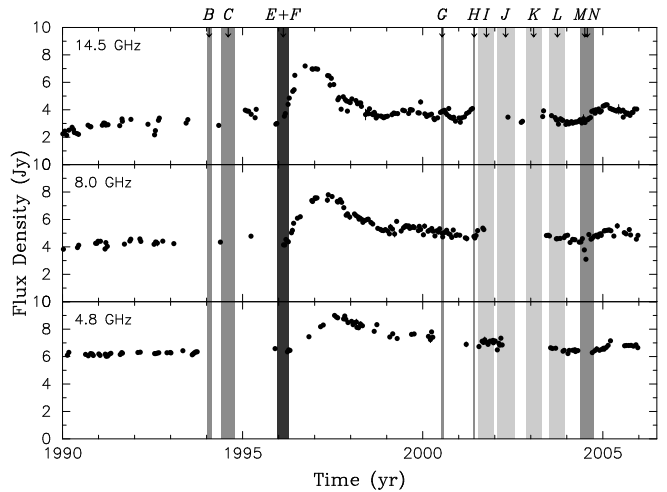}
   \caption{University of Michigan Radio Astronomy Observatory light curves of 3C\,111 at 4.8\,GHz, 8\,GHz, and
14.5\,GHz. The shaded
areas indicate the ejection epochs of the individually labeled jet components
as discussed in Sect.~\ref{sect:vlbaresults}. 
The lightest shading corresponds to minor ejections of the relatively weak
components I,J,K,L with flux densities $S$ below $0.2$\,Jy, medium shading corresponds to
components B,C,G,H,M,N with $0.2{\rm Jy} < S < 0.6 {\rm Jy}$ and the darkest shading to
components E and F with $S > 0.6 {\rm Jy}$.
}
\label{fig:3c111_umrao}
\end{figure}

\begin{figure}[t]
   \centering
   \includegraphics[clip,angle=-90,width=\linewidth]{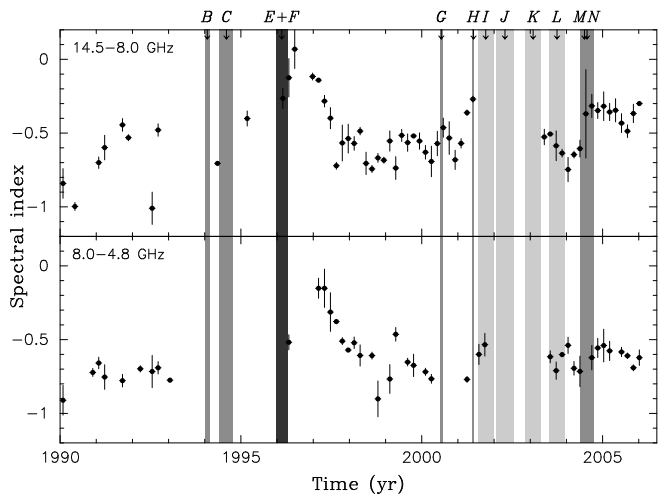}
   \caption{Spectral-index curves of 3C\,111 between 14.5\,GHz and 8\,GHz (top), and
8\,GHz and 4.8\,GHz (bottom) from the UMRAO monitoring program. The shaded
areas are the same as in Fig.~\ref{fig:3c111_umrao}.
}
\label{fig:3c111_spit}
\end{figure}

\subsection{The 1996 Radio Outburst of 3C\,111}

\begin{figure*}[t!]
\centering
\includegraphics[clip, angle=-90,width=\linewidth]{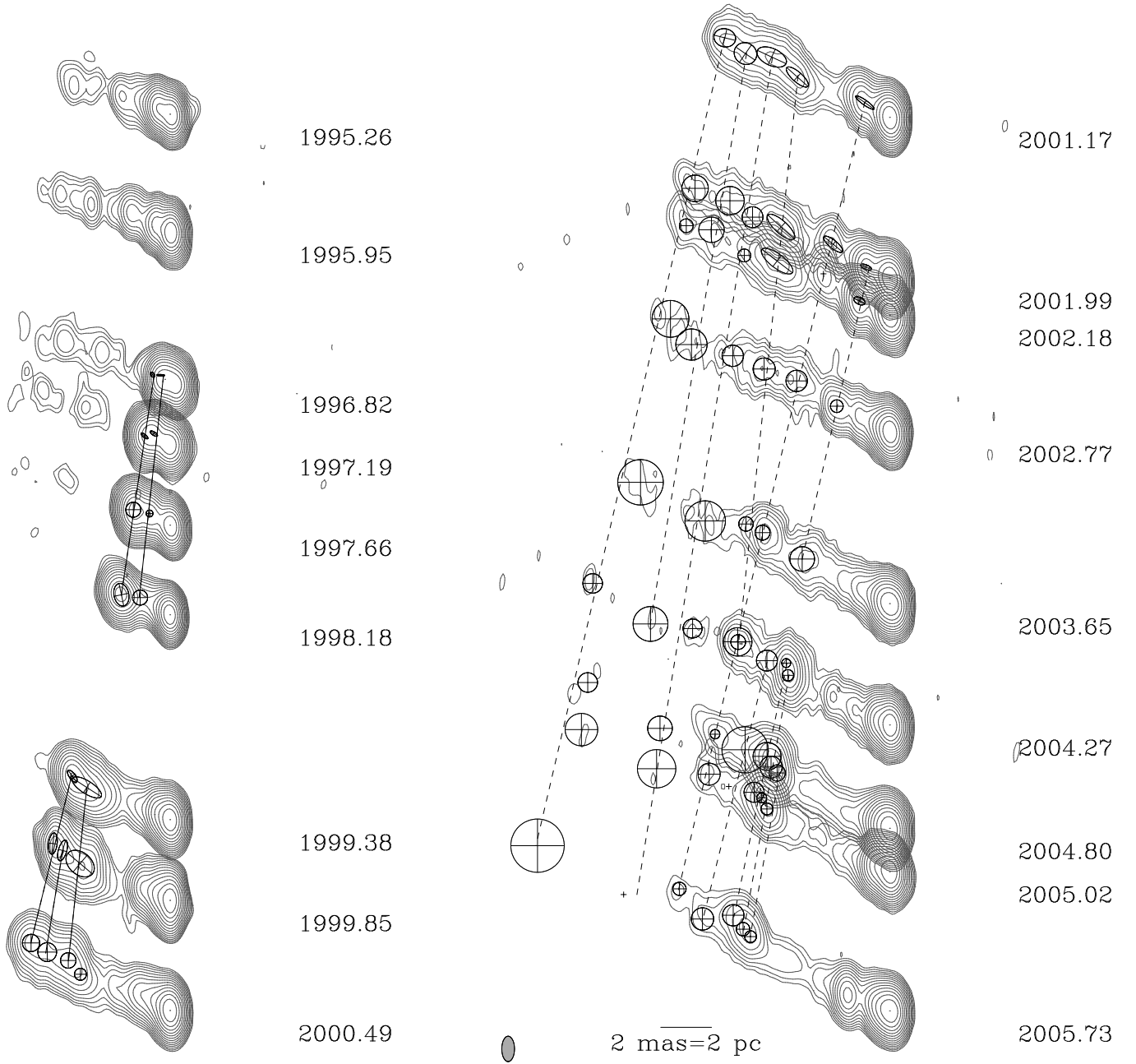}
\caption{
Naturally weighted images of the parsec-scale jet of 3C\,111 from the 2\,cm VLBA monitoring.
A common restoring beam of $(0.5 \times 1.0)$\,mas at P.A. $0^\circ$ was used.
The total recovered flux density in each image, the rms noise, and the lowest contours for each image are
given in Table~\ref{tab:3c111_journal}.
Contours increase logarithmically by a factor of $2$. Only components E, G, H and their
corresponding trailing components are indicated by circles enclosing a cross.
\label{fig:3c111_allepochs}
}
\end{figure*}

A strong flux density outburst of 3C\,111 occurred in
1996, which was first visible in the mm band and some months later at 
lower radio frequencies. This outburst was first detected
at 90\,GHz with the IRAM interferometer at Plateau de Bure
in January 1996 with flux densities greater than $10$\,Jy \citep{Ale98},
at 37\,GHz in March 1996, and at 22\,GHz in August 1996 with the
Mets\"ahovi radio observatory \citep{Ter04}.
Figure~\ref{fig:3c111_umrao} shows the single-dish radio light curves of
3C\,111 at 4.8\,GHz, 8\,GHz, and
14.5\,GHz obtained from the UMRAO radio-flux-density monitoring program \citep{All03}.
These data show that from early 1996 on, the radio-flux density of 3C\,111
was rising at 14.5\,GHz, reaching its maximum  in late 1996. At the two lower
frequencies, the flux-density maximum was reached at subsequent later times,
in mid 1997 at 8\,GHz and in late 1997 at 4.8\,GHz.
The profile of the outburst in the flux-density vs. time domain shows a narrow, high-amplitude
peak between early 1996 and late 1997, which is almost symmetric. After late 1997, a slower-decreasing
component dominates the light curves, most clearly visible at 14.5\,GHz.

Figure~\ref{fig:3c111_umrao} shows that the flare propagated through
the spectrum as qualitatively expected by standard jet theory \citep[e.g.,][]{Mar85}: high-frequency
radio emission comes from the most compact regions of the jet, the emission
peak shifts to lower frequencies as a newly ejected jet component travels down
the jet and becomes optically thin. The peak flux density shifted with frequency
at about $10$\,GHz\,yr$^{-1}$.

The evolution of the spectral index, $\alpha$ ($S\sim\nu^\alpha$), for (14.5/8.0)\,GHz and (8.0/4.8)\,GHz is shown in
Fig.~\ref{fig:3c111_spit}. Before 1996, the sampling was too sparse to derive the change
of the spectral index in the (14.5/8.0)\,GHz band. Between, 8.0\,GHz and 4.8\,GHz,
the spectral index was approximately $-0.7$ during the pre-1996 period. The radio flux-density outburst
in 1996 corresponded to a subsequent flattening of the spectrum with a maximum 
spectral index, $\alpha \sim 0$,  
in the (14.5/8.0)\,GHz band reached in mid 1996. In the post-outburst period
between 1998 and 2004, $\alpha$ was typically in the range $-0.5$ to  $-0.7$ between 14.5\,GHz
and 8.0\,GHz and slightly steeper ($-0.7$ to $-0.9$) in the (8.0/4.8)\,GHz band. The overall
steeper spectral index at lower frequencies can be understood as the contribution of
optically thin
large-scale emission from the radio lobes of 3C\,111 to these single-dish light curves.

\subsection{VLBA Monitoring Results}
\label{sect:vlbaresults}

Figure~\ref{fig:3c111_allepochs} shows the variable parsec-scale structure of 3C\,111
at 18 different epochs
of VLBA observations between 1995.26 and 2005.73.
The variable source structure can be described by a classical one-sided
core-jet morphology in the first two epochs with typical velocities of the
outward moving jet components of about $1.4$ to $1.7$\,mas\,yr$^{-1}$ corresponding
to about $5$\,$c$. In 1996.82 a new jet component, even brighter than the core,
dominated the source structure. By 1997.19, this new component was even brighter ($\sim 3.4$\,Jy)
and in the following epochs
it traveled along the jet while it became gradually more stretched out along the jet-ridge line.

\begin{figure}
   \centering
   \includegraphics[clip,angle=-90,width=\linewidth]{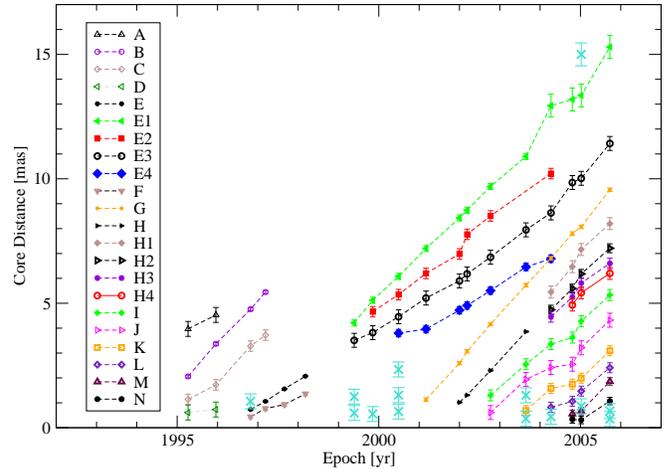}
   \caption{Core separation of model-fit components vs. time.
Crosses represent components which could not be cross identified over the epochs.
Position uncertainties have been estimated from the internal deviations of the data
from linear motions for each component. The uncertainties for the position of E\,1 have been determined
separately for the pre-2004 and post-2004 epochs because of the partial resolution of
this component after 2004. Uncertainties for components with less than
three epochs were estimated from other components at similar positions in the jet and with similar flux
densities.
}

              \label{fig:3c111_rad}
\end{figure}

\begin{figure}
   \centering
   \includegraphics[clip,angle=-90,width=\linewidth]{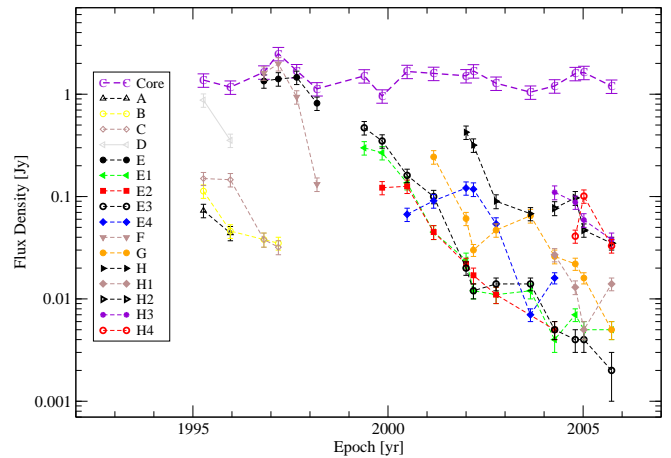}
   \caption{Flux-density evolution of the core and the jet components with time.
For clarity, only components ejected before 2001.50 are shown. 
The flux densities of E\,1, E\,2, E\,3, and E\,4 were added for the post-1999 epochs and
a flux-density weighted effective position was calculated 
to display the flux-density evolution of the blended feature
that would be visible at lower resolution.
Note that components E\,4 and G are blended
in epoch 2004.27 and that the flux density of E\,4 may be overestimated for this epoch.
}

              \label{fig:3c111_fluxes}
\end{figure}

\paragraph{Model Fitting:}
In Fig.~\ref{fig:3c111_rad}, the radial distance of the various model fit components from the core
is shown as a function of time.
The component identification was based on a comparison of the positions and
flux densities, and a linear regression of the distances from the core as a function
of time was used to determine the kinematics. 
The derived component velocities are tabulated
in Table~\ref{tab:3c111_kinematics}.
The early outer jet components (A, B, C, D) of the 1995.26 epoch can be
traced over two to four epochs before their flux densities fall below the detection
threshold (compare Fig.~\ref{fig:3c111_fluxes}).
In late 1996 and early 1997, the source structure was dominated by the emission of the core
and the newly formed jet components E and F, with E being the leading component. The two components
traveled outwards with a mean apparent velocity of $(1.00 \pm 0.02)$\,mas\,yr$^{-1}$
and $(0.64 \pm 0.07)$\,mas\,yr$^{-1}$, respectively.
Before mid 1997, component F was substantially brighter than component E but after that, its flux
density dropped steeply. F was not detected at any epoch later than 1998.18, while E was still
about $800$\,mJy at that time. The light curves of E and F reproduce qualitatively the
two-component shape of the flux-density outburst in Fig.~\ref{fig:3c111_umrao}
with component F being responsible for the narrower and higher-amplitude peak between early 1996 and
late 1997 and component E dominating the slower-decreasing tail of the outburst after late 1997
(compare Fig.~\ref{fig:3c111_fluxes} and discussion below).
In the following epochs, E split into four distinct components (E\,1, E\,2, E\,3, E\,4)
at distances of 3.5\,mas to 4.5\,mas from the core.
E\,2, E\,3, and E\,4 all moved at subsequently slower speeds than E\,1, resulting in an elongated
morphological structure of the associated emission complex.

In later epochs, new components have been ejected from the core
into the jet. The two strongest components 
(G,H) can be traced through the following eight and nine monitoring epochs,
respectively. Component H split into three individual components in 2004.27 and a fourth associated
component was seen from 2004.80 on. In the following, we refer to the components E\,1, and H\,1 as the
``leading components'' and to E\,2, E\,3, E\,4, H\,2, H\,3, and H\,4 as the ``trailing components''
of E and H, respectively.

\begin{figure}[t!]
   \centering
   \includegraphics[clip,angle=-90,width=\linewidth]{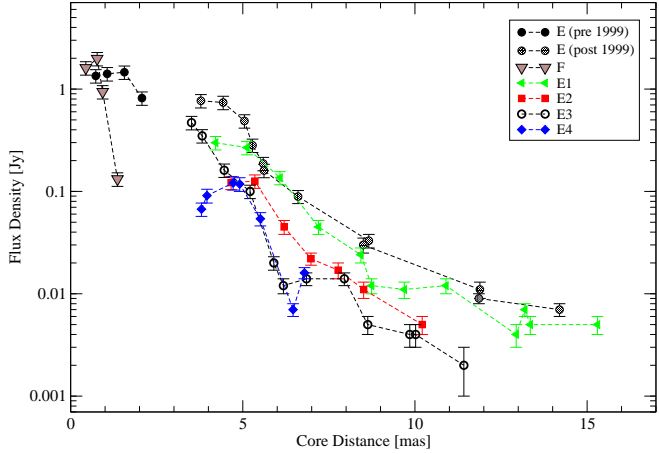}
   \caption{Flux-density evolution of component E and its trailing components vs. the distance 
traveled from the core.
Note that component E\,4 is blended with component G
in epoch 2004.27 and that its flux density may be overestimated for this epoch.
}
\label{fig:3c111_fluxes2a}
\end{figure}
\begin{figure}[h!]
   \centering
   \includegraphics[clip,angle=-90,width=\linewidth]{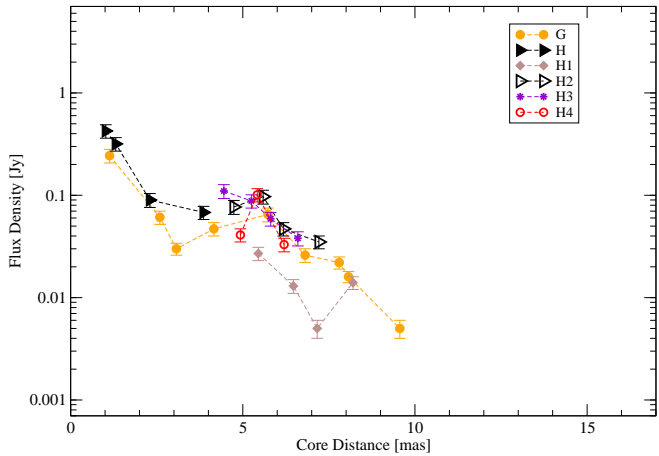}
   \caption{Flux-density evolution of component G and H and its trailing components vs. distance traveled from the core.
}
\label{fig:3c111_fluxes2b}
\end{figure}

For the pc-scale jet of 3C\,111, the
ejection epochs of the individual jet components can be determined 
from the linear regression by
back-extrapolating the component trajectories to the core. In Fig.~\ref{fig:3c111_umrao}, these ejection
epochs and the associated uncertainties are indicated as shaded areas. It is apparent that the ejection of
the components E and F coincides with the onset of the major flux-density outburst in 1996 described above.
The following major component ejections (G, H, and the combined M/N event) all have direct counterparts
in local maxima of the radio light curve, especially at 14.5\,GHz.
Figure~\ref{fig:3c111_spit} shows that all these ejection epochs coincided with local maxima of the
spectral index in the 14.5/8.0\,GHz band.
Between 2002 and 2004, a number of minor
component ejections took place but the regression-fit quality (due to the neareby components, the low
flux densities and the small time baseline) only moderatly constrains the ejection epochs. In addition,
the time sampling of UMRAO observations in this time range is relatively poor, in particular from mid 2001 to
mid 2003.

\begin{figure}[t!]
   \centering
   \includegraphics[clip,angle=-90,width=\linewidth]{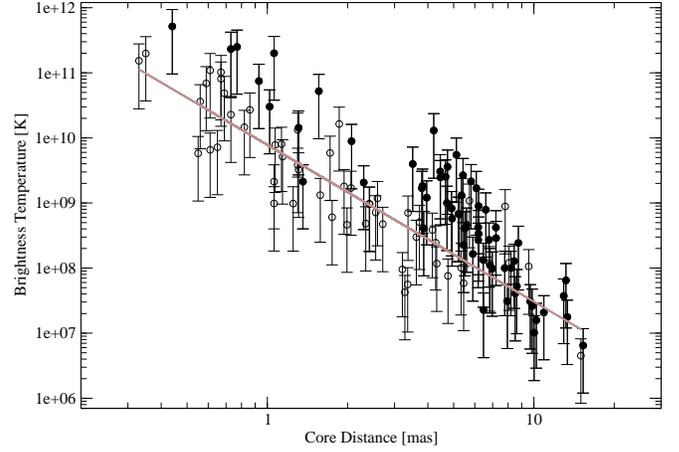}
   \caption{Brightness temperatures of model-fit components as a function of their distance to the core. The brightness temperatures
of components belonging to the E--, F-- and H--components are indicated by filled black circles. The solid line represents a least-squares
fit to all but the E--, F-- and H--components. The slope of the regression curve is $-2.4 \pm 0.2$.
}
\label{fig:3c111_tb}
\end{figure}
\begin{figure}[h!]
   \centering
   \includegraphics[clip,angle=-90,width=\linewidth]{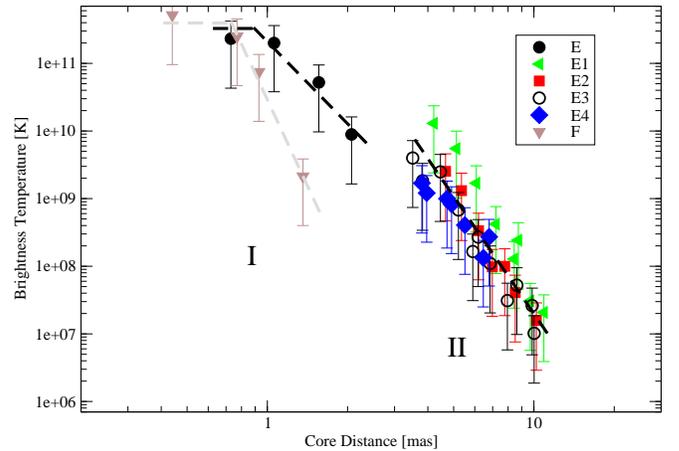}
   \caption{Brightness temperatures of component E, its leading and trailing components,
and component F
as a function of their distance to the core. The two regimes of brightness-temperature gradient
discussed in the text are indicated with dashed lines.
}
\label{fig:3c111_tb_E}
\end{figure}

\paragraph{Flux Density Evolution:}
Figure~\ref{fig:3c111_fluxes} shows the brightness evolution of the core and the jet components
that have been ejected prior to 2001.5. 
Apparently, the trailing components E\,4 and H\,4 appeared first in a rising
state, i.e., they first increased in flux density before they became fainter in later epochs. Component
F showed an extraordinary steep decrease in brightness in 1997--1998.
In Fig.~\ref{fig:3c111_fluxes2a} and Fig.~\ref{fig:3c111_fluxes2b}, the flux-density evolution of the components
E, G, and H and the associated leading and trailing components
are shown with distance traveled from the core, respectively.
The ejecta first rose in flux density within the inner 1\,mas from the core, then they showed a decline
about almost three orders of magnitude in the following decade, exhibiting a plateau or broad local maximum
in 1998--2000 at a distance from 2--4\,mas from the core.
Component H and its leading and trailing components exhibited a similar behavior although on about an order of magnitude
lower flux-density levels and at slightly further downstream, 4--6\,mas from the core.
Component G, in spite of the fact that it does not appear to have split into leading and trailing
components like E and H, did exhibit a pronounced flux density maximum after 2002, as well,
about 6\,mas from the core.

\paragraph{The T$_{\rm b}$ gradient along the jet:}
Following \citet{Kad04,Kad05}, the power-law index $s$, which describes the brightness
temperature gradient via $T_{\rm b} \propto r^{s}$, can be parametrized as
\begin{equation}\label{eq:s}
s\,=\,l\,+\,n\,+\,b\,(1\,-\,\alpha)
\end{equation}
where $l$, $n$ and $b$ are the power law indices that describe the
gradients of jet diameter $d \propto r^l$, particle density $n_e \propto r ^n$, and
magnetic field $B \propto r^b$ with distance $r$ from the core, respectively.
Therefore, measuring the brightness temperature gradient  
provides a method to constrain the critical physical properties 
along the jet and abrupt changes in the $T_{\rm b}$-gradient
can highlight regions in the jet where the density, magnetic field, or jet diameter
change rapidly. 

Figure~\ref{fig:3c111_tb} shows the brightness temperatures of all jet components
in the parsec-scale jet of 3C\,111 at 2\,cm wavelength between 1995 and 2005 as
a function of their core distance. In general, the brightness temperature of all components
decreased as the components traveled outwards but an approximation with a simple power
law does not yield a good fit to the full data set ($\chi_{\rm red}^2=1.8$, 115 degrees of freedom [d.o.f.]).
Visual inspection of Fig.~\ref{fig:3c111_tb} shows that this is due to the E--, F-- and H-- components
and their leading and trailing components, respectively.
This behavior is different than
expected for a straight and stable jet geometry in which the power-law dependences of the particle
density, the magnetic field strength and the jet diameter on the core distance predicts
that the brightness temperature along the jet can be described with a
well-defined power-law index $s$. Most extragalactic parsec-scale jets
which do not show pronounced curvature, show a power-law decrease with
increasing distance from the core and power-law indices typically around $-2.5$
\citep{Kad05}.
In fact, excluding the E--, F-- and H--components from the fit yields a statistically better
result ($\chi_{\rm red}^2=1.3$, 52 d.o.f.) and a gradient of $-2.4 \pm 0.2$.
In Sect.~\ref{sect:smaller1mas} and Sect.~\ref{sect:3-5mas}, we discuss possible 
physical reasons for the different behavior and nature of these components.
The measured relation between component sizes and distance along the jet is
affected by a large degree of
scatter and does not provide independent information from the flux
density and
brightness temperature plots described above. Therefore we do not
show plots of component size versus jet distance.

The brightness-temperature gradient of component E was first flat or inverted immediately after
the creation of this new component within approximately 1\,mas from the core
and then reached steep values of $-2.5$ to $-2.8$ (regime I; compare Fig.~\ref{fig:3c111_tb_E})
through 1997 when the component
traveled from 1\,mas to 2\,mas.
Between 2\,mas and 4\,mas, the determination of the brightness-temperature gradient requires
an identification of component E with either component E\,1 or E\,3 (see below).
Independently of this identification, the brightness-temperature
gradient eventually changed to very steep values ($s<-5$) beyond 5\,mas from the core (regime II).
Component F began its very rapid decline in brightness temperature at a very small distance
from the core ($< 0.7$\,mas) with an extremely steep $T_{\rm b}$-gradient ($s <-8$).

\begin{figure}[t!]
   \centering
   \includegraphics[trim=1.3cm 0cm 0cm 0cm,width=\linewidth]{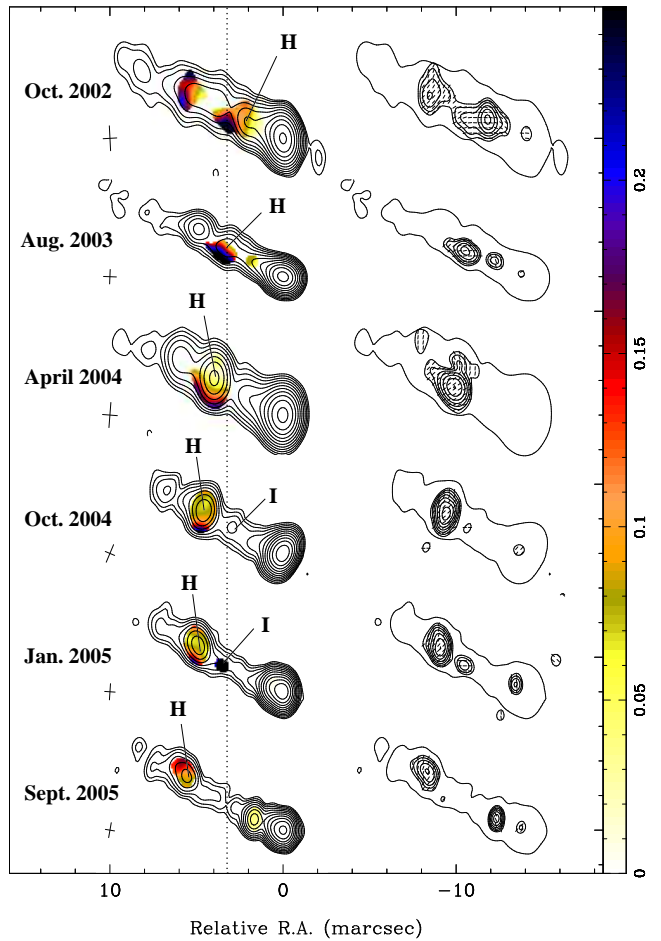}
   \caption{Naturally weighted images of the linear-polarization structure of 3C\,111
between 2002 and 2005.
The restoring-beam dimensions and orientations
for each epoch are indicated by a cross to the left of each
Stokes-I image.
Stokes I contours start at 1\,mJy/beam and increase by factors of 
$2$.  Fractional polarization is over-plotted on the Stokes-I contours in
color.
To the right of the Stokes-I images are the polarization
intensity contours starting at 1\,mJy/beam and increasing by factors 
$\sqrt{2}$.  The polarization contours are over-plotted with
tick-marks representing the electric vector position angle.  A single
Stokes-I 1\,mJy contour
surrounds the polarization image to show registration.
The dotted line marks the distance of 3.3\,mas from the core 
where the most pronounced changes of the polarization properties take place 
(see text). 
}
\label{fig:3c111_pol}
\end{figure}

\paragraph{Linear Polarization:}
From 1995 to 2002, 2\,cm-Survey observations were done in
left circular polarization only, so no linear-polarization
information can be derived from these data.
MOJAVE observations (from 2002 on) are done in full-polarimetric mode.
Figure~\ref{fig:3c111_pol}  shows our polarization
data through September 2005. 

In October of 2002, component H was about 2.5 mas from the base of
the jet and showed a fractional polarization of 5\% to 10\% increasing
towards the downstream side of the component.  The electric
vector position angle (EVPA) displayed by the component was approximately
aligned with the jet.
In this epoch, the jet material just downstream of
component H 
at $\sim 3.3$\,mas from the base of the jet
was more highly polarized, exceeding 20\% fractional
polarization on the jet's southern side, and the EVPA of the
polarization turned to be about 45$^\circ$ to the main jet
direction.

By August of 2003, component H had entered a region approximately $3.3$\,mas from the core
and its observed polarization was now similar to
the emission in this same region observed in the previous epoch.
The observed fractional polarization of H now climbed sharply to values in
excess of 20\% toward the jet's southern side
while there was no detectable polarization from the northern
side of H.  The observed EVPA of H had rotated further to be approximately
60$^\circ$ to the main jet direction. However, the EVPA 
on the southern-most side of H was 
approximately perpendicular to the jet. 
 
After component H passed through this region (epochs April 2004
through September 2005), it split into a number of subcomponents as
described earlier, and its polarization gradually became more
uniform. Consistent fractional polarization of 5\% to 10\% was approached
with the electric vectors approximately perpendicular to the local
jet direction.
 
The much weaker component, I,
developed polarization very similar to H as it passed through
the same region, about 3.3 mas from the core, with fractional
polarization exceeding 20\% toward the southern side of the
jet with an EVPA at approximately 45$^\circ$ to the main jet axis.  This
is also the same region of the jet in which component E had broken
up into a number of sub-components.  In future epochs, we
will have the opportunity to follow component K as it passes
through this same region.

\section{Discussion}
\label{sect:discussion}
In this section, we discuss the 
aftermath of the major outburst in 3C\,111 in 1996
and the following component ejections through 2005.
We organize the subsections of our discussion
according to the downstream distance from the VLBI core
where we observe the effect of interest.

\subsection{Within $1$\,pc: Forward and Reverse Structures}
\label{sect:smaller1mas}
Numerical simulations (\citealt{Alo03}) show that an abrupt
perturbation of the fluid density at the jet injection point
during a short time propagates downstream, evolves spreading
asymmetrically along the jet and finally splits into two distinct
regions. Both of these two regions have enhanced energy density
with respect to the underlying jet, and they emit synchrotron
radiation. The leading (forward shock) and the following region (reverse shock)
have higher and lower Lorentz factors, respectively, 
than the underlying jet. 
Thus, they should separate with time as they propagate downstream in
the jet. 

Component F matches the description of a backward moving wave
associated with the major injection into the jet of 3C\,111 after
the flux-density outburst of 1996. It follows the trail of
component E but at a lower speed. If 
component F is identified with a reverse shock and component E with a forward
shock, it is possible then to compute the size of the shocked
region \citep{Per07}. In
1996.82 and 1997.19, E and F were both very bright and separated by
only $\sim 0.3$\,pc in projected distance. During these two
epochs, F was 300\,mJy to 500\,mJy brighter than the leading
component E. Following \citet{Alo03}, a backward shock can be
brighter than a forward shock if the latter is beamed in a cone
smaller than the viewing angle due to its larger speed. We have
examined the Doppler factors of components E and F for the range of
possible viewing angles (see Appendix~\ref{sect:inclination}) 
and the measured
velocities and conclude that this alone cannot explain
the brightness difference between component E and F
because the difference in apparent speed is not large enough. 
\citet{Jor05} point out that backward
shocks can be brighter than forward shocks as long as the
disturbance is prolonged and there is a continuous supply of
particles entering from the underlying jet through the shock region. Within
half a year, between 1997.19 and 1997.66, F lost about half of
its brightness. This extraordinarily fast dimming of the backward
shock can be caused by the lack of input of particles from behind,
i.e., a lower plasma ejection rate after the primary injection
possibly due to a depletion of the inner accretion disk \citep{Mar02,Mar06} which feeds
the plasma injection.

Component F can also be interpreted as a rarefaction propagating
backwards in the reference frame of the ejected blob of gas. A
rarefaction is produced when the blob is overpressured
with respect to the jet, as this overpressure causes the front to
accelerate in the jet, thus leaving a rarefied region between the
head of the blob (forward shock) and its rear part, which is still
slower (it moves with the injection velocity). In this case, the
emission in component F could be associated to the denser and
overpressured gas in the blob which has still not been rarefied.
This gas would cease to emit as soon as it reaches the
rarefied region, which may also explain the sudden decrease in
brightness of this component. An extended discussion on the nature
of component F and the evolution of its brightness will be given
in \citet{Per07}.


\subsection{Between $2$\,pc and $4$\,pc: Expansion and Acceleration}
  It is not \emph{a-priori} clear with which post-split-up
component the original feature E should be identified after 1999.
A natural identification would be the leading component E1 but that
requires an acceleration of this component (see Fig.~4) from
$\beta_{\rm app,E}=3.26\pm0.07$ to $\beta_{\rm app,E1}=5.5\pm0.1$
between 1998.18 and 1999.38. This may be interpreted in terms of
an expansion of the jet in a rarefied medium. Taking an angle to
the line of sight of $19^\circ$ (see
Appendix~\ref{sect:inclination}), the component would be accelerated
from $\beta=0.956$ ($\gamma\sim3.4$) to $\beta=0.995$
($\gamma\sim10.3$). The increase of velocity is less at smaller
viewing angles. 

An alternative model for the acceleration and brightening of component E 
would be a change of the jet inclination to the line of sight from about 24$^\circ$ to about 11$^\circ$ 
at this location in the jet as observed in the case of the quasar 3C\,279 \citep{Hom03}.
However, we see no significant change in jet position angle in the sky which would be expected
to accompany such a large change in jet direction. Moreover, subsequent components, 
particularly G, do not show the same kind of large acceleration in this region.

Direct identification of component E with component E1 is not
straightforward in the frame of expansion, as component E1 in
epoch 1999.38 was smaller than component E in 1998.18 (see
Table~2). However, component E3 in epoch 1999.38 is larger than
component E in 1998.18. We can interpret this as component E
including components E1 and E3 (and maybe E4). These components
would be indistinguishable in our observations before 1999.38.
In fact, \cite{Jor05}
monitored 3C\,111 between 1998 and 2001 with
the VLBA at 43GHz. They find an emission complex, that can be
identified with our component E, that gradually stretches out as
it travels from $\sim$ 2 mas from the core in 1998 to roughly between $5$\,mas and $8$\,mas 
from the core in 2001. Their leading component C1 can be
identified with our component E1, their component c2 with E2 and
their c1 with E3.
At their higher angular resolution, Jorstad et al.\nocite{Jor05} can separate
components C\,1 and  c\,1 already in early 1998.
In agreement with our analysis at 15\,GHz, they
detect c2 (E2) about a year after they detect c1 (E3).
They do not detect a component
corresponding to E4 but this may be an effect of partially resolving out the
jet structure at their higher observing frequency, particularly in later epochs.
It is
further interesting to note that the observed speeds at both frequencies agree well. 
For E1(C1),  $\mu_{\rm app, 2\,cm} = 1.69 \pm 0.04$\,mas\,yr$^{-1}$ at 15 GHz and 
$\mu_{\rm app, 7\,mm} = 1.77 \pm 0.06$\,mas\,yr$^{-1}$ at 43 GHz; 
for E2(c2), $\mu_{\rm app, 2\,cm} =  1.29 \pm 0.06$\,mas\,yr$^{-1}$ at 15 GHz and 
$\mu_{\rm app, 7\,mm} = 1.23 \pm 0.04$\,mas\,yr$^{-1}$ at 43 GHz;
and for E3(c1), 
$\mu_{\rm app, 2\,cm} = 1.22 \pm 0.05$\,mas\,yr$^{-1}$ at 15 GHz and 
$\mu_{\rm app, 7\,mm} = 1.07 \pm 0.02$\,mas\,yr$^{-1}$ at 43 GHz.
The discrepancy in the speeds measured for E3 and c1 seems to be due to a
slight acceleration of E3 after 2002. A fit to the 15 GHz data of
E3 between 1999 and 2002 alone yields a slower speed of $\sim$ 1.0
mas\,yr$^{-1}$ similar to the speed of c1 in the same time period at
43\,GHz. In their work, Jorstad et al.\nocite{Jor05} do not report acceleration of components
from 2\,mas to 4\,mas. However, this is likely due to the fact
that their observations started in early 1998, thus missing the
first observations of component E presented in this paper, when
its speed has been measured to be smaller.

\subsection{Between $2$\,pc and $6$\,pc: Recollimation of the Jet}
\label{sect:recollimation}
Inspection of Fig.~9 shows that the back-extrapolation of the
brightness temperature of component E from regime II to regime I is at
least two orders of magnitude too high if this
extrapolation is based on the gradient given by E1. 
The low brightness temperature of component E in regime I cannot be
explained by opacity effects because the radio-light curve in
Fig.~1 shows that the source was optically thin 
from 1997 on. Moreover, if we identify component E with E1, it is
Doppler-deboosted from epoch 1998.18 to 1999.38 due to the
acceleration and a relatively large viewing angle; thus, we are
not able to explain the increase in brightness temperature in
terms of Doppler boosting. 
However, compact sub-components may have larger brightness temperatures, so that the
$T_{\rm b}$ values plotted in Fig.~9 for E in regime I inward of about $3$\,mas
may represent lower limits for
compact components already embedded in the unresolved structure.

Not only E/E1 but also components G and H show an increase in
total flux density several milliarcseconds downstream. Compared to
E/E1, these somewhat weaker components exhibit their flux-density
maxima at somewhat larger distances from the core (compare Fig.~6
and Fig.~7). 
This can be explained if the gas in the components travels through a
mild standing shock in a recollimation region. 
This effect has been 
observed in numerical simulations of parsec \citep{Gom97}
and kiloparsec scale jets \citep{Per07b}.
The
material in the components is expected to be overpressured with
respect to its environment, thus expanding into it. After the
initial expansion, the components become underpressured with
respect to the underlying flow. The resulting recollimation leads
to the formation of a shock, whose strength depends on the initial
degree of overpressure of the material in the component. This
process explains the increase in flux density and brightness
temperature as due to compression of the gas in the recollimation.
In Figs.~6 and 7, we see that the flux density of component E
increases closer to the core
than for component G and H, which is consistent with the former being slower
than the latter, thus recollimating earlier \citep[see][]{Per07b}.
It also
explains why we see a significant acceleration only in the faster
expanding, brighter component E/E1.

Finally, after this mild recollimation, the fluid becomes
overpressured with respect to its environment, thus further
expanding and accelerating downstream.

\subsection{Near $3$\,pc: The Role of the External Medium}
\label{sect:polarization}
The polarization behavior of components H and I can be
understood in terms of
an interaction between the jet and the external medium at a
distance of ~3.3\,mas ($= 3.3$\,pc) in the jet.
Assuming no Faraday rotation, the 
EVPA of component H within approximately 3.3\,mas from the core 
indicates a transverse magnetic field order
as might be expected for a transverse shock propagating down the
jet. The change in the 
fractional polarization, its north-south gradient, and the rotation of the EVPA
suggest that a contact surface persists at the
southern boundary of the jet beam at a
distance of approximately 3.3\,mas downstream the jet core.
If the bulk jet material flows faster than
the flow at the southern boundary, the magnetic field is stretched
through shear.
Our overall picture then is of an originally transverse
shock interacting with the jet on the southern side of the
jet at ~ 3.3\,mas from the core.  The interaction changes
the component's magnetic field through some combination
of oblique shock and differential flow resulting in a
magnetic field approximately parallel to the jet axis in
the later epochs.
No strong shock is needed at this location
in the jet
but this region may be identified with the recollimation
region (see Sect.~\ref{sect:recollimation}) at about the
same position in the jet).
In this picture, the jet-medium interaction
may form an effective nozzle which accelerates the jet on
one edge relative to the other.

An alternative explanation for the observed polarization structure and dynamics of 3C\,111
can be found by considering an inhomogeneous external Faraday screen. Such a screen could
produce the observed differential rotation of the EVPA while a component travels
through a given region along the jet. \cite{Zav02} observed 3C\,111 with the VLBA
and produced a Faraday rotation-measure map between 8\,GHz and 15\,GHz. They find strong
Faraday rotation, $\sim 730$\,rad\,m$^{-2}$, at the same distance from the core
(3.3\,mas) where our observations show the swing of the EVPA
of the component H and steeply decreasing Faraday rotation further downstream.
However, we note that $730$\,rad\,m$^{-2}$ translates to 17$^\circ$ of rotation at 15\,GHz
which alone is not enough to explain the change in EVPA that we observe while
component H travels through this region. On the other hand, the steep decrease
of the Faraday rotation measured up- and downstream of this region by \citet{Zav02} again 
agrees with a change of the external gas density at this point, which in turn
may be identified with the pressure gradient responsible for the
component expansion and accelleration.

A combination of inhomogeneous Faraday rotation and an interaction between the jet plasma and its ambient medium appears most likely to explain our observations of the varying linear polarization structure;
however, both explanations point to the role of the external medium, either through a discrete interaction or a rapid decrease in external gas pressure, in shaping the  jet flow downstream of this location.

\subsection{Between $3$\,pc and $5$\,pc: Formation of Trailing Components}
\label{sect:3-5mas}
The components E2, E3, and E4 can be interpreted as
trailing components forming in the wake of the leading E1 which is
identical with the original component E. This scenario is
attractive because the basic concept of trailing components as
introduced by \citet{Agu01} predicts the formation of
trailing features in the wake of the initial perturbation in the
jet flow. Such a behaviour has first been found both associated with bright
sub- and superluminal jet components in Centaurus A and 3C 120 \citep{Tin01,Gom01}.
\citet{Jor05} report trailing components in four additional sources
(3C\,273, 3C\,345, CTA\,102 and 3C\,454.3) and in 3C\,111 (see below).

The interaction of the external medium
with a strong shock pinches the surface of the jet, leading to the
production of pinch-body mode Kelvin- Helmholtz instabilities: the
trailing features. Hence, a single strong superluminal shock
ejection from the jet nozzle may lead to the production of a
multiple set of emission features through this mechanism. The
trailing features have a characteristic set of properties, which
make them recognizable with high resolution VLBI: 
they form in the wake of strong components
instead of being ejected from the core of VLBI jets, they are
related to oblique shocks, they are always slower than the leading
feature, and (if the underlying jet has a certain opening angle)
they should be generated with a wide range of apparent speeds
(from almost stationary near the core to superluminal further
downstream). Moreover, \citet{Agu01} showed that the separation
between the trailing components increases downstream due to their
motion down a pressure gradient.

All this is in agreement with what we observe in the trailing
components of E\,1 and with our interpretation of an expansion of
the jet in a density decreasing ambient medium. 
For the time range covered by their observations (1998.23 to 2001.28),
\citet{Jor05} also identified the
{trailing phenomenology} in this source.

The
north-south gradients detected in the linear-polarization emission
in the region where the trailing features are formed, is in agreement with an
oblique shock structure. The steep brightness-temperature
gradients of the trailing components  indicate that the particle
and magnetic field density associated with these components evolve
in a different way compared to the ``normal" jet flow. These
shocked regions may be more overpressured with respect to
their environment, making them expand rapidly. This fast
expansion implies a larger positive value of $l$, which, however,
is compensated by an even larger (negative) value of $n$ and $b
(1-\alpha)$ in equation~\ref{eq:s}, resulting in a very steep
brightness temperature gradient (regime II).

Pinching modes of the Kelvin-Helmholtz instability were shown to
couple to the trailing components observed in the simulations in
\citet{Agu01}. In the case of components E2-E4, the distance
between them ranges from 0.7-0.8 mas at the first epochs in which
they are observed, to almost 2.0 mas in the latest epochs. Taking
into account that: a) their FWHM is of the same order (see Table
2); b) that these wavelengths have to be corrected for 
geometrical and relativistic effects, resulting in a maximum
intrinsic wavelength of $\sim 0.7\,\rm{mas}$, and c) that the size
of the components can be of the order or smaller than the jet
radius (\citealt{Per07a}), this implies coupling of the pinching to
wavelengths of the order or smaller than the jet radius.
\citet{Per07} have shown that resonant Kelvin-Helmholtz
instabilities associated to high-order body modes appear in
sheared jets at these wavelengths. These modes have larger growth
rates than low-order body modes or surface modes, and their growth
brings the jet to a final quasi-steady state in which it remains
well-collimated and generates a hot shear-layer which shields the
core of the jet from the ambient medium. Interestingly, the jet in
3C~111 is known to be well-collimated up to kiloparsec scales.
Further research in this direction is needed in order to check the
influence of the resonant modes in the long term evolution of this
jet.

A by-product of the interpretation of these components as
Kelvin-Helmholtz instabilities is the fact that it allows us to
put constraints to the velocity of the jet. We can regard the wave
speed as the minimum speed of the jet flow, as KH modes have an
upper limit in their wave speeds that is precisely the velocity of
the flow in which they propagate \citep{Per06}. The upper limit is given by the
speed of E1, interpreted as a shock wave, that has to be thus
faster than the underlying flow. In this picture, we would have
the structure E1 moving with Lorentz factor $\gamma\sim 8.3$
through a jet with Lorentz factor $8.3>\gamma_j\geq 4.6$ in the
accelerated region (post 1999.38), where the lower limit is given
by the Lorentz factor of component E2, the fastest of the three
trailing components identified here.

\section{Summary and Conclusions}
\label{sect:conclusions}
In this paper, we have investigated the parsec-scale
jet kinematics and the interaction of the jet with its ambient
medium in the broad-line radio galaxy 3C\,111. Our analysis has
demonstrated that a variety of processes influence the jet
dynamics in this source: a plasma injection into the jet beam
associated with a major flux-density outburst leads to the
formation of multiple shocks that travel at different speeds
downstream and interact with each other and with the ambient
medium. The primary perturbation causes the formation of a forward
and a backward shock (or rarefaction). The latter fades away so
fast that is likely to remain undetected in minor ejections. A
separate work by \citet{Per07} focuses on the nature and
characteristics of these initial components. Several parsecs
downstream, the jet plasma enters a region of rapidly decreasing
external pressure, expands into the jet ambient medium and
accelerates. In the following, the plasma gets recollimated and
trailing features are formed in the wake of the leading component.

A particularly interesting aspect of the source 3C\,111 in the 
light of this and other recent works is that it is one of 
the very rare non-blazar gamma-ray bright AGN. Besides Centaurus\,A
\citep{Sre99} and the possible identification of NGC\,6251 
with the EGRET source 3EG\,J1621+8203 \citep{Muk02}, 3C\,111
is the only AGN whose jet-system is inclined at a relatively large
angle to the line of sight and that has a reliable EGRET identification:
\citet{Sgu05} reconsidered the possible identification
of the EGRET source 3EG\,J0416+3650 with 3C\,111, which was first suggested by
\citet{Har99} but considered unlikely because of the poor
positional coincidence. Very recently, R.\,C.~Hartman \& M.~Kadler (in prep.) 
found that 3EG\,J0416+3650
is composed out of at least two distinct components.
One of them is the dominant source above 1\,GeV and is in excellent positional agreement with the
location of 3C\,111. Compared to blazars, the large inclination angle
and the relatively small distance of 3C\,111 allow us to resolve
structures along the jet that are as small as parsecs in deprojection
and which would be heavily blended with adjacent features in blazar jets.
As demonstrated in this paper, VLBA observations of 3C\,111 probe 
a variety of physically
different regions in a relativistic extragalactic jet such as a compact
core, superluminal jet components, recollimation shocks and regions
of interaction between the jet and its surrounding medium, which are all
possible sites of gamma-ray production. 
From early 2008 on, the gamma-ray satellite GLAST \citep{Lot07} is going to monitor
the sky. 
If 
detected by GLAST, 3C\,111 may become a key source in the quest for an understanding
of the origin of gamma-rays from extragalactic jets. 
In addition, the combination of GLAST and VLBA
data with spectral data at intermediate
wavelengths (optical, IR, X-ray) may allow a better determination of 
jet parameters and relativistic beaming effects than in most blazars
because of the higher linear resolution offered by this nearby and only weakly projected
jet system.

Our observations of 3C\,111 are qualitatively in remarkable agreement
with numerical relativistic hydrodynamic structural and emission
simulations of jets such as the ones presented by \cite{Agu01} and
\cite{Alo03}. Further progress is being made in the transition from
two-dimensional to three-dimensional simulations of relativistic jets
and in the development of new methods considering magnetic fields
\citep[e.g.,]{Lei05,Miz07,Roc08}, the equation of state for relativistic gases \citep{Per07b}, and radiative processes \citep[e.g.,][and Mimica et al. in preparation]{Mim04,Mim07}. But so far neither observational data nor simulations have reached an adequate level of detail and completeness in order to allow us a quantitative direct comparison of
numerical models and observed relativistic jet structure and evolution. 
In particular, it is not 
feasible
today to fit iteratively the parameters of relativistic magneto-hydro-dynamical (RMHD) jet simulations to match the brightness distribution observed for any individual source. 
The main reasons for this are a) the immense computational power required to conduct a realistic (i.e., sufficiently detailed) modern 3D jet simulation and b) the  highly non-linear nature of RMHD plasmas and their evolution. Simulation results depend critically on the starting conditions like the exact velocity, composition, and transversal structure of the flow, the structure and strength of the magnetic field and the jet environment. Future development of computational power will allow us to use larger resolutions to decrease the numerical viscosities, and to implement nonlinear and microphysics processes into simulations. VLBA observations are capable of putting hard quantitative constraints on the input parameters for RMHD jet simulations if they are densely sampled over several years. Polarimetric observations at multiple radio frequencies may allow the
effects of jet-intrinsic magnetic-field variations and external Faraday-screen inhomogenities or temporal variations to be disentangled. Such data at 15\,GHz are on the way, e.g., as part of the next phase of the MOJAVE program, in which rapidly evolving sources like 3C\,111 are being observed every two months.

\acknowledgments
We would like to acknowledge the support of the rest of the MOJAVE Team, who have contributed
to the data used in this paper, in particular we would like to thank Christian Fromm for his help
with the production of the figures for the paper.
We thank Dharam Vir Lal and Silke Britzen for their careful reading of the manuscript and their comments.
We also thank the referee for his
very constructive suggestions, which have helped to improve this paper.
We are grateful to Greg Taylor, who provided complementary 2\,cm  VLBA data for an additional
epoch.
The MOJAVE project is supported under National Science Foundation
grant 0406923-AST.
The
Very Long Baseline Array is operated by the National Radio Astronomy Observatory, a
facility of the National Science Foundation operated under cooperative agreement by Associated Universities, Inc.
UMRAO is partially supported  by a series of grants from the NSF, most
recently AST-0607523, and by funds
from the University of Michigan. 
MK has been supported in part by a Fellowship of the
International Max Planck Research School for Radio and Infrared Astronomy and in part by
an appointment to the NASA Postdoctoral Program at the Goddard Space Flight Center,
administered by Oak Ridge Associated Universities through a contract with NASA.
MP acknowledges support in part by the Spanish \emph{Direcci\'on
General de Ense\~nanza Superior} under grant AYA2004-08067-C03-01
and in part by a postdoctoral fellowship of the \emph{Generalitat Valenciana} (\emph{Beca Postdoctoral
d'Excel$\cdot$l\`encia}).
YYK is a Research Fellow of the Alexander von Humboldt
Foundation and was supported in part by the Russian Foundation
for Basic Research (project 05-02-17377).
DCH was supported by grants from Research Corporation and the National Science Foundation (AST-0707693)
IA has been supported in part by an I3P contract with the Spanish Consejo Superior de Investigaciones Cient\'{i}ficas and in part by a contract with the German Max-Planck-Institut f\"ur Radioastronomie (through the ENIGMA network, contract HPRN-CT-2002-00321), which 
were partially funded by the EU.

\appendix
\section{The Jet Inclination Angle}
\label{sect:inclination}
The 1996 radio outburst of 3C\,111 puts strong constraints on the angle to the line of sight for this source,
if one assumes that a similarly bright component as E has been ejected in the counterjet, as well. Due to differential
Doppler boosting, the flux density ratio between the jet- and counter-jet emission is
\begin{equation}
\frac{S_{\rm J}}{S_{\rm CJ}} = \left(\frac{1+\beta \cos{\theta}}{1-\beta \cos{\theta}}\right)^{2-\alpha} \quad .
\end{equation}
Thus, for a given jet to counter-jet ratio $x = \frac{S_{\rm J}}{S_{\rm CJ}}$
\begin{equation}
\beta \cos{\theta}  = \frac{x-1}{x+1} \quad .
\end{equation}
With $\alpha = 0.3$, $S_{\rm J} = 3.4$\,Jy (components E and F in 1997.19), $S_{\rm CJ} < 10$\,mJy and $\beta < 1$, $\theta < 21^\circ$.
For a realistic jet speed of, e.g., $\beta = 0.956$ ($\gamma = 3.4$), the angle to the line
of sight is: $\theta = 19^\circ$.
An estimate close to this value
can be derived from the variability Doppler factor measured by \citet{Lae99} and the apparent superluminal jet speed.
As outlined in detail in \citet{Coh07}, this leads to a value of $\theta \sim 15^\circ$.

It is important to note that this calculation implicitly assumes symmetry between the jet and counter-jet, which
in projection does not have to be the case if the counter-jet is covered by an obscuring torus
as it is well-established for systems at larger inclination angles \citep[e.g., NGC\,1052: see ][]{Kad04}. Indeed,
Faraday rotation measurements towards the 3C\,111 pc-scale jet (\citealt{Zav02}; see also Sect.~\ref{sect:polarization})
and X-ray spectral observations
(\citealt{Lew05}) suggest substantial amounts
of obscuring material. Free-free absorption could also substantially lower the counter-jet radio emission and allow for
larger jet angles to the line of sight.

An independent lower limit on the inclination angle of $\theta > 21^\circ$ was given by \cite{Lew05} assuming that the
deprojected size of the largescale 3C\,111 double-lobe structure is smaller than $500 h^{-1}$\,kpc. This discrepancy
implies that either 3C\,111 is unusually large or there is a misalignement between the large-scale jet-axis and the parsec-scale
jet axis inclination to the line of sight, although the projected position angles of the large-scale jet ($63^\circ$) and the
parsec-scale jet ($\sim 65^\circ$) are almost the same.

\clearpage
\section{Image-Plane vs. $(u,v)$-plane Model Fitting}

\begin{figure*}[h]
   \centering
   \includegraphics[width=\linewidth]{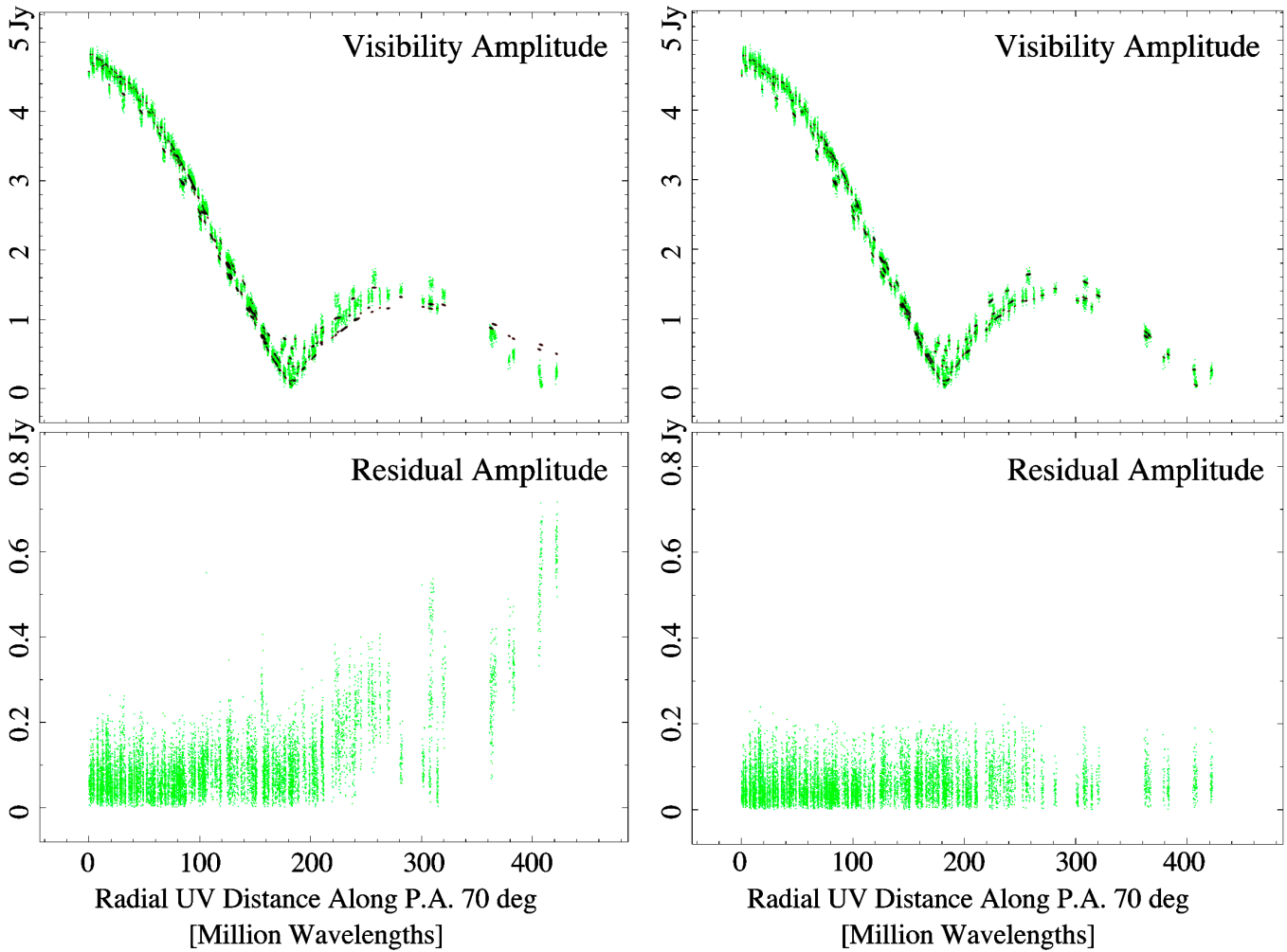}
   \caption{Visibility amplitudes projected radially along
P.A. 70$^\circ$ for the 1996.82 observation of 3C\,111.
The double-peak indicates the presence of a bright structure
within less than a milliarcsecond of the core.
The top left panel shows a model (black) which was fitted to the data (green) consisting of
one model component for the newly ejected jet feature and the residuals
of this model are shown in the bottom left panel. Up to about $600$\,mJy of
correlated flux density is missing from the model. The right panels show the same data
fitted by a model consisting of two components (corresponding to E and F).
}
\label{fig:projplots}
\end{figure*}

It is interesting to compare our results from this very detailed analysis
of one individual object with the results of the kinematical-survey
analysis of \cite{Kel04}, who investigated the speeds of 110 extragalactic jets, including
3C\,111, based on the VLBA 2\,cm Survey data between 1994 and 2001.
\cite{Kel04} made the component identification in the image plane and
represented the evolving structure of component E and its trailing features
by only one component 
whereas, in this work we distinguish the sub-components E, F, E1, E2, E3, and E4. 
Formally, the speed of $(4.9 \pm 0.2) c$ found by \citet{Kel04} is in good agreement
with the speed of our leading component, E1, so 
the much simpler model derived from image-plane analysis represents the
fastest moving structure. The acceleration (with respect to the ejecta's
smaller velocity prior to mid 1999), as well as the additional components that
we interpret as a backward shock and trailing jet features become visible
only after a more complicated model fitting of the data in the
$(u,v)$-domain. The necessarily less-complex model used in a survey
analysis like the one conducted by \cite{Kel04} is only part of the reason for
this discrepancy. Image-plane analysis makes it very difficult to interpret
a moving feature that changes its structure in a complex way and that
has no clear persistent brightness maxima. In addition, $(u,v)$-plane
fitting in general achieves higher angular resolution so that the
two bright, but closely separated components E and F could not be distinguished
in in early to mid 1996. Figure~\ref{fig:projplots} demonstrates that even in
October 1996 when E and F are separated by only 0.3\,mas and are located within
1\,mas from the core, a one-component model
clearly fails to represent this compact structure.

\bigskip

\clearpage
\begin{table}
      \caption[]{Journal of VLBA 2cm Survey observations of 3C\,111 analyzed in this study.
         \label{tab:3c111_journal}}
\centering
\resizebox{0.6\columnwidth}{!}{%
            \begin{tabular}{@{}clcccc@{}}
\noalign{\smallskip}
\hline
\hline
\noalign{\smallskip}
Epoch  & Code &  $S_{\rm tot}$ & rms & $C^{\rm \star}$ & $m^{\rm \dag}$\\
       &      &     [Jy] & [mJy/beam] & [mJy/beam] & [\%]\\
            \noalign{\smallskip}
\hline
\noalign{\smallskip}
1995.27 & BK\,016 &  $2.6$ & $1.8$ & $6.5$ \\
\noalign{\smallskip}
1995.96$^{\rm a}$ & BK\,037A &  $1.8$ & $1.1$ & $3.4$ \\
\noalign{\smallskip}
1996.82 & BK\,37D & $4.8$ & $0.5$ & $2.1$  \\
\noalign{\smallskip}
1997.19 & BK\,048 &   $6.0$ & $0.9$ & $4.9$ \\
\noalign{\smallskip}
1997.66 & BK\,052A &  $4.1$ & $1.6$ & $6.4$ \\
\noalign{\smallskip}
1998.18 & BK\,052B &  $2.0$ & $2.1$  & $5.5$ \\
\noalign{\smallskip}
1999.39 & BK\,068A &  $2.6$ & $0.6$ & $1.9$ \\
\noalign{\smallskip}
1999.85 & BK\,068C &  $2.4$ & $1.2$  & $4.1$ \\
\noalign{\smallskip}
2000.49 & BT\,051  &  $2.6$ & $0.5$ & $1.4$ & 1.5\\
\noalign{\smallskip}
2001.17 & BK\,068E &  $2.1$ & $0.3$ & $1.3$ \\
\noalign{\smallskip}
2002.00 & BR\,077D &  $2.2$ & $0.3$ & $1.1$ \\
\noalign{\smallskip}
2002.19 & BR\,077I &  $2.2$ & $0.2$ & $0.9$ \\
\noalign{\smallskip}
2002.77 & BL\,111C &  $1.6$ &  $0.5$ & $1.7$ & 0.6 \\
\noalign{\smallskip}
2003.65 & BL\,111J &  $2.1$ & $0.3$ & $1.2$ & 0.3 \\
\noalign{\smallskip}
2004.27$^{\rm a}$ & BL\,111L &  $1.9$ & $0.4$ & $1.4$ & 0.7\\
\noalign{\smallskip}
2004.80$^{\rm b}$ & BL\,111P &  $2.7$ & $0.3$ & $1.0$ & 0.7\\
\noalign{\smallskip}
2005.02 & BL\,123A &  $3.0$ & $0.5$ & $1.8$ & 0.5\\
\noalign{\smallskip}
2005.73 & BL\,123O &  $2.4$ & $0.3$ & $1.2$ & 0.7\\
\noalign{\smallskip}
\hline
\multicolumn{6}{l}{\footnotesize $^{\rm \star}$ Lowest contour in Fig.~\ref{fig:3c111_allepochs}} \\
\multicolumn{6}{l}{\footnotesize $^{\rm \dag}$ Degree of polarization (see Fig.~\ref{fig:3c111_pol})} \\
\multicolumn{6}{l}{\footnotesize $^{\rm a}$ No data from antenna at Mauna Kea}\\
\multicolumn{6}{l}{\footnotesize $^{\rm b}$ No data from antenna at St.Croix}
\end{tabular}
}
\end{table}

\clearpage
\LongTables 
\begin{deluxetable}{ccccccc}
\tablewidth{0pt}
\tablecaption{Model fit parameters
\label{tab:3c111-modelfits} }
\tablehead{
\vspace{2.0ex} \\
\colhead{ID }           & \colhead{Flux Density [mJy]}      &
\colhead{Radius [mas]}          & \colhead{P.A.$^{\rm a}$ [$^\circ$]}  &
\colhead{FWHM [mas]}          & \colhead{ratio}    &
\colhead{$\phi$[$^\circ$]}
}  
\startdata
\vspace{2.0ex} \\
\noalign{\smallskip}
\multicolumn{7}{c}{\it \hrulefill 1995.27  \hrulefill} \\
\noalign{\smallskip}
0	&	1371.04	$\pm$	205.66	&	0.00	&	24.00	&	0.33	$\pm$	0.60	&	0.43	&	56.22	\\
D	&	876.06	$\pm$	131.41	&	0.61	$\pm$	0.30	&	61.18	&	0.21	$\pm$	0.54	&	1.00	&	--	\\
C	&	149.59	$\pm$	22.44	&	1.14	$\pm$	0.21	&	60.80	&	0.41	$\pm$	0.64	&	1.00	&	--	\\
B	&	112.85	$\pm$	16.93	&	2.06	$\pm$	0.29	&	69.41	&	0.62	$\pm$	0.79	&	1.00	&	--	\\
A	&	73.30	$\pm$	10.99	&	3.97	$\pm$	0.30	&	70.69	&	0.00	$\pm$	0.50	&	1.00	&	--	\\
\noalign{\smallskip}
\multicolumn{7}{c}{ \it \hrulefill 1995.96 \hrulefill } \\
\noalign{\smallskip}
0	&	1173.64	$\pm$	176.05	&	0.00	&	--	&	0.35	$\pm$	0.12	&	0.59	&	47.42	\\
D	&	353.66	$\pm$	53.05	&	0.73	$\pm$	0.30	&	61.85	&	0.30	$\pm$	0.12	&	1$^{\rm b}$	&	--	\\
C	&	145.60	$\pm$	21.84	&	1.72	$\pm$	0.21	&	62.73	&	0.38	$\pm$	0.13	&	1$^{\rm b}$	&	--	\\
B	&	46.15	$\pm$	6.92	&	3.37	$\pm$	0.29	&	70.25	&	0.61	$\pm$	0.16	&	1$^{\rm b}$	&	--	\\
A	&	43.54	$\pm$	6.53	&	4.53	$\pm$	0.30	&	69.99	&	-$^{\rm c}$			&	1$^{\rm b}$	&	--	\\
\noalign{\smallskip}
\multicolumn{7}{c}{ \it \hrulefill 1996.82 \hrulefill } \\
\noalign{\smallskip}
0	&	1645.03	$\pm$	246.75	&	0.00	&	--	&	0.26	$\pm$	0.11	&	0.17	&	60.63	\\
F	&	1608.94	$\pm$	241.34	&	0.44	$\pm$	0.07	&	54.71	&	0.31	$\pm$	0.12	&	0.19	&	87.46	\\
E	&	1344.91	$\pm$	201.74	&	0.73	$\pm$	0.02	&	67.83	&	0.23	$\pm$	0.11	&	0.61	&	35.24	\\
X1	&	121.29	$\pm$	18.19	&	1.06	$\pm$	0.3	&	61.59	&	0.33	$\pm$	0.12	&	1$^{\rm b}$	&	--	\\
C	&	38.24	$\pm$	5.74	&	3.28	$\pm$	0.21	&	66.12	&	0.88	$\pm$	0.20	&	1$^{\rm b}$	&	--	\\
B	&	38.04	$\pm$	5.71	&	4.76	$\pm$	0.06	&	68.08	&	0.66	$\pm$	0.16	&	1$^{\rm b}$	&	--	\\
\noalign{\smallskip}
\multicolumn{7}{c}{ \it \hrulefill 1997.19 \hrulefill } \\
\noalign{\smallskip}
0	&	2490.20	$\pm$	373.53	&	0.00	&	--	&	0.38	$\pm$	0.13	&	0.26	&	62.36	\\
F	&	1980.30	$\pm$	297.05	&	0.77	$\pm$	0.07	&	55.51	&	0.34	$\pm$	0.12	&	0.39	&	57.52	\\
E	&	1410.05	$\pm$	211.51	&	1.06	$\pm$	0.02	&	71.09	&	0.34	$\pm$	0.12	&	0.35	&	51.00	\\
C	&	32.15	$\pm$	4.82	&	3.72	$\pm$	0.21	&	65.48	&	0.23	$\pm$	0.11	&	1$^{\rm b}$	&	--	\\
B	&	35.26	$\pm$	5.29	&	5.45	$\pm$	0.06	&	67.95	&	0.72	$\pm$	0.18	&	1$^{\rm b}$	&	--	\\
\noalign{\smallskip}
\multicolumn{7}{c}{ \it \hrulefill 1997.66 \hrulefill } \\
\noalign{\smallskip}
0	&	1697.09	$\pm$	254.56	&	0.00	&	--	&	0.51	$\pm$	0.14	&	0$^{\rm c}$	&	59.71	\\
F	&	940.20	$\pm$	141.03	&	0.93	$\pm$	0.07	&	60.22	&	0.27	$\pm$	0.11	&	1$^{\rm b}$	&	--	\\
E	&	1461.49	$\pm$	219.22	&	1.56	$\pm$	0.02	&	67.18	&	0.40	$\pm$	0.13	&	1$^{\rm b}$	&	--	\\
\noalign{\smallskip}
\multicolumn{7}{c}{ \it \hrulefill 1998.18 \hrulefill } \\
\noalign{\smallskip}
0	&	1128.63	$\pm$	169.29	&	0.00		&	--	&	0.82	$\pm$	0.19	&	0.47	&	6.27	\\
F	&	131.71	$\pm$	19.76	&	1.36	$\pm$	0.07	&	59.83	&	0.59	$\pm$	0.16	&	1$^{\rm b}$	&	--	\\
E	&	815.70	$\pm$	122.35	&	2.07	$\pm$	0.02	&	67.53	&	0.90	$\pm$	0.21	&	0.65	&	11.93	\\
\noalign{\smallskip}
\multicolumn{7}{c}{ \it \hrulefill 1999.39 \hrulefill } \\
\noalign{\smallskip}
0	&	1509.63	$\pm$	226.44	&	0.00	&	--	&	0.38	$\pm$	0.13	&	0$^{\rm c}$	&	64.00	\\
X3	&	225.98	$\pm$	33.90	&	0.59	$\pm$	0.30	&	62.46	&	0.14	$\pm$	0.10	&	1$^{\rm b}$	&	--	\\
X2	&	54.38	$\pm$	8.16	&	1.25	$\pm$	0.30	&	65.08	&	0.56	$\pm$	0.15	&	1$^{\rm b}$	&	--	\\
E3	&	470.42	$\pm$	70.56	&	3.51	$\pm$	0.28	&	68.80	&	1.34	$\pm$	0.29	&	0.37	&	57.41	\\
E1	&	298.52	$\pm$	44.78	&	4.21	$\pm$	0.12	&	66.02	&	0.56	$\pm$	0.15	&	0.42	&	35.72	\\
\noalign{\smallskip}
\multicolumn{7}{c}{ \it \hrulefill 1999.85 \hrulefill } \\
\noalign{\smallskip}
0	&	960.22	$\pm$	144.03	&	0.00	&	--	&	0.78	$\pm$	0.19	&	0.41	&	-4.69	\\
X4	&	743.68	$\pm$	111.55	&	0.55	$\pm$	0.30	&	59.07	&	1.13	$\pm$	0.25	&	0.58	&	61.59	\\
E3	&	349.03	$\pm$	52.35	&	3.82	$\pm$	0.28	&	67.78	&	1.26	$\pm$	0.27	&	0.69	&	50.04	\\
E2	&	122.27	$\pm$	18.34	&	4.66	$\pm$	0.21	&	65.12	&	0.84	$\pm$	0.20	&	0.39	&	-16.32	\\
E1	&	267.68	$\pm$	40.15	&	5.12	$\pm$	0.12	&	64.17	&	0.84	$\pm$	0.19	&	0.40	&	-7.05	\\
\noalign{\smallskip}
\multicolumn{7}{c}{ \it \hrulefill 2000.49 \hrulefill } \\
\noalign{\smallskip}
0	&	1668.98	$\pm$	250.35	&	0.00	&	--	&	0.33	$\pm$	0.12	&	0.11	&	66.21	\\
X7	&	249.65	$\pm$	37.45	&	0.65	$\pm$	0.30	&	56.02	&	0.45	$\pm$	0.13	&	1$^{\rm b}$	&	--	\\
X6	&	133.82	$\pm$	20.07	&	1.31	$\pm$	0.30	&	59.94	&	0.48	$\pm$	0.14	&	1$^{\rm b}$	&	--	\\
X5	&	43.07	$\pm$	6.46	&	2.34	$\pm$	0.30	&	66.11	&	0.71	$\pm$	0.17	&	1$^{\rm b}$	&	--	\\
E4	&	66.54	$\pm$	9.98	&	3.80	$\pm$	0.15	&	68.04	&	0.47	$\pm$	0.14	&	1$^{\rm b}$	&	--	\\
E3	&	160.55	$\pm$	24.08	&	4.46	$\pm$	0.28	&	63.86	&	0.61	$\pm$	0.16	&	1$^{\rm b}$	&	--	\\
E2	&	125.58	$\pm$	18.84	&	5.35	$\pm$	0.21	&	64.62	&	0.74	$\pm$	0.18	&	1$^{\rm b}$	&	--	\\
E1	&	135.77	$\pm$	20.37	&	6.08	$\pm$	0.12	&	64.13	&	0.68	$\pm$	0.17	&	1$^{\rm b}$	&	--	\\
\noalign{\smallskip}
\multicolumn{7}{c}{ \it \hrulefill 2001.17 \hrulefill } \\
\noalign{\smallskip}
0	&	1596.68	$\pm$	239.50	&	0.00	&	--	&	0.44	$\pm$	0.13	&	0$^{\rm c}$	&	61.41	\\
G	&	243.56	$\pm$	36.53	&	1.13	$\pm$	0.07	&	58.96	&	0.88	$\pm$	0.20	&	0.22	&	54.85	\\
E4	&	91.23	$\pm$	13.68	&	3.96	$\pm$	0.15	&	66.28	&	1.11	$\pm$	0.24	&	0.35	&	48.34	\\
E3	&	99.76	$\pm$	14.96	&	5.21	$\pm$	0.28	&	62.80	&	1.24	$\pm$	0.27	&	0.54	&	65.22	\\
E2	&	45.34	$\pm$	6.80	&	6.20	$\pm$	0.21	&	66.06	&	0.91	$\pm$	0.21	&	0.92	&	55.42	\\
E1	&	45.18	$\pm$	6.78	&	7.20	$\pm$	0.12	&	64.16	&	0.86	$\pm$	0.20	&	0.83	&	75.97	\\
\noalign{\smallskip}
\multicolumn{7}{c}{ \it \hrulefill 2002.00 \hrulefill } \\
\noalign{\smallskip}
0	&	1514.03	$\pm$	227.10	&	0.00	&	--	&	0.49	$\pm$	0.14	&	0.09	&	61.18	\\
H	&	424.80	$\pm$	63.72	&	1.02	$\pm$	0.02	&	64.67	&	0.46	$\pm$	0.14	&	0.39	&	66.86	\\
G	&	60.57	$\pm$	9.09	&	2.59	$\pm$	0.07	&	58.88	&	0.94	$\pm$	0.21	&	0.33	&	50.61	\\
E4	&	121.10	$\pm$	18.16	&	4.72	$\pm$	0.15	&	64.66	&	1.40	$\pm$	0.30	&	0.35	&	47.76	\\
E3	&	19.67	$\pm$	2.95	&	5.90	$\pm$	0.28	&	65.92	&	0.82	$\pm$	0.19	&	1$^{\rm b}$	&	--	\\
E2	&	21.62	$\pm$	3.24	&	6.98	$\pm$	0.21	&	64.03	&	1.12	$\pm$	0.25	&	1$^{\rm b}$	&	--	\\
E1	&	24.46	$\pm$	3.67	&	8.43	$\pm$	0.12	&	65.14	&	1.04	$\pm$	0.23	&	1$^{\rm b}$	&	--	\\
\noalign{\smallskip}
\multicolumn{7}{c}{ \it \hrulefill 2002.19 \hrulefill } \\
\noalign{\smallskip}
0	&	1688.54	$\pm$	253.28	&	0.00	&	--	&	0.56	$\pm$	0.15	&	0.09	&	63.02	\\
H	&	317.77	$\pm$	47.66	&	1.31	$\pm$	0.02	&	64.52	&	0.48	$\pm$	0.14	&	0.56	&	65.20	\\
G	&	29.58	$\pm$	4.44	&	3.07	$\pm$	0.07	&	57.88	&	-$^{\rm c}$			&	1$^{\rm b}$	&	--	\\
E4	&	118.14	$\pm$	17.72	&	4.91	$\pm$	0.15	&	64.22	&	1.51	$\pm$	0.32	&	0.36	&	52.47	\\
E3	&	12.28	$\pm$	1.84	&	6.18	$\pm$	0.28	&	67.59	&	0.51	$\pm$	0.14	&	1$^{\rm b}$	&	--	\\
E2	&	17.46	$\pm$	2.62	&	7.77	$\pm$	0.21	&	64.31	&	1.00	$\pm$	0.22	&	1$^{\rm b}$	&	--	\\
E1	&	12.35	$\pm$	1.85	&	8.74	$\pm$	0.12	&	66.18	&	0.54	$\pm$	0.15	&	1$^{\rm b}$	&	--	\\
\noalign{\smallskip}
\multicolumn{7}{c}{ \it \hrulefill 2002.77 \hrulefill } \\
\noalign{\smallskip}
0	&	1280.01	$\pm$	192.00	&	0.00	&	--	&	0.47	$\pm$	0.14	&	0$^{\rm c}$	&	62.23	\\
J	&	100.40	$\pm$	15.06	&	0.61	$\pm$	0.28	&	68.62	&	0.30	$\pm$	0.12	&	1$^{\rm b}$	&	--	\\
I	&	78.58	$\pm$	11.79	&	1.30	$\pm$	0.22	&	61.52	&	0.34	$\pm$	0.12	&	1$^{\rm b}$	&	--	\\
H	&	90.33	$\pm$	13.55	&	2.30	$\pm$	0.02	&	64.39	&	0.50	$\pm$	0.14	&	1$^{\rm b}$	&	--	\\
G	&	47.49	$\pm$	7.12	&	4.16	$\pm$	0.07	&	61.40	&	0.84	$\pm$	0.20	&	1$^{\rm b}$	&	--	\\
E4	&	53.64	$\pm$	8.05	&	5.51	$\pm$	0.15	&	63.49	&	0.87	$\pm$	0.20	&	1$^{\rm b}$	&	--	\\
E3	&	13.63	$\pm$	2.04	&	6.85	$\pm$	0.28	&	64.21	&	0.84	$\pm$	0.20	&	1$^{\rm b}$	&	--	\\
\vspace{2.0ex} \\
E2	&	10.84	$\pm$	1.63	&	8.51	$\pm$	0.21	&	66.27	&	1.23	$\pm$	0.27	&	1$^{\rm b}$	&	--	\\
E1	&	11.31	$\pm$	1.70	&	9.69	$\pm$	0.12	&	62.75	&	1.45	$\pm$	0.31	&	1$^{\rm b}$	&	--	\\
\noalign{\smallskip}
\multicolumn{7}{c}{ \it \hrulefill 2003.65 \hrulefill } \\
\noalign{\smallskip}
0	&	1046.67	$\pm$	157.00	&	0.00	&	--	&	0.32	$\pm$	0.12	&	0.20	&	74.73	\\
X9	&	576.24	$\pm$	86.44	&	0.36	$\pm$	0.30	&	68.30	&	0.09	$\pm$	0.10	&	1$^{\rm b}$	&	--	\\
K	&	177.19	$\pm$	26.58	&	0.69	$\pm$	0.20	&	66.00	&	0.14	$\pm$	0.10	&	1$^{\rm b}$	&	--	\\
X8	&	66.75	$\pm$	10.01	&	1.30	$\pm$	0.30	&	62.77	&	0.17	$\pm$	0.11	&	1$^{\rm b}$	&	--	\\
J	&	34.80	$\pm$	5.22	&	1.94	$\pm$	0.28	&	63.40	&	0.33	$\pm$	0.12	&	1$^{\rm b}$	&	--	\\
I	&	21.58	$\pm$	3.24	&	2.71	$\pm$	0.22	&	62.07	&	0.51	$\pm$	0.14	&	1$^{\rm b}$	&	--	\\
H	&	67.76	$\pm$	10.16	&	3.86	$\pm$	0.02	&	62.83	&	0.97	$\pm$	0.22	&	1$^{\rm b}$	&	--	\\
G	&	64.55	$\pm$	9.68	&	5.73	$\pm$	0.07	&	60.73	&	0.59	$\pm$	0.15	&	1$^{\rm b}$	&	--	\\
E4	&	7.48	$\pm$	1.12	&	6.46	$\pm$	0.15	&	60.88	&	0.57	$\pm$	0.15	&	1$^{\rm b}$	&	--	\\
E3	&	13.78	$\pm$	2.07	&	7.95	$\pm$	0.28	&	65.78	&	1.59	$\pm$	0.33	&	1$^{\rm b}$	&	--	\\
E1	&	11.72	$\pm$	1.76	&	10.90	$\pm$	0.12	&	63.95	&	1.79	$\pm$	0.37	&	1$^{\rm b}$	&	--	\\
\noalign{\smallskip}
\multicolumn{7}{c}{ \it \hrulefill 2004.27 \hrulefill } \\
\noalign{\smallskip}
0	&	1204.12	$\pm$	180.62	&	0.00	&	--	&	0.31	$\pm$	0.12	&	0.14	&	70.05	\\
X10	&	270.09	$\pm$	40.51	&	0.44	$\pm$	0.30	&	68.51	&	-$^{\rm c}$			&	1$^{\rm b}$	&	--	\\
L	&	105.50	$\pm$	15.83	&	0.82	$\pm$	0.20	&	69.58	&	0.20	$\pm$	0.11	&	1$^{\rm b}$	&	--	\\
K	&	38.46	$\pm$	5.77	&	1.58	$\pm$	0.20	&	66.52	&	0.41	$\pm$	0.13	&	1$^{\rm b}$	&	--	\\
J	&	26.88	$\pm$	4.03	&	2.42	$\pm$	0.28	&	63.12	&	0.40	$\pm$	0.13	&	1$^{\rm b}$	&	--	\\
I	&	11.76	$\pm$	1.76	&	3.36	$\pm$	0.22	&	65.95	&	1.09	$\pm$	0.24	&	1$^{\rm b}$	&	--	\\
H3	&	110.03	$\pm$	16.50	&	4.45	$\pm$	0.21	&	63.91	&	0.45	$\pm$	0.14	&	1$^{\rm b}$	&	--	\\
H2	&	76.83	$\pm$	11.53	&	4.74	$\pm$	0.18	&	59.02	&	0.35	$\pm$	0.12	&	1$^{\rm b}$	&	--	\\
H1	&	26.80	$\pm$	4.02	&	5.45	$\pm$	0.24	&	62.26	&	0.82	$\pm$	0.19	&	1$^{\rm b}$	&	--	\\
G	&	26.20	$\pm$	3.93	&	6.81	$\pm$	0.07	&	61.25	&	1.12	$\pm$	0.25	&	1$^{\rm b}$	&	--	\\
E4	&	15.80	$\pm$	2.37	&	6.79	$\pm$	0.15	&	61.23	&	0.58	$\pm$	0.15	&	1$^{\rm b}$	&	--	\\
E3	&	5.15	$\pm$	0.77	&	8.63	$\pm$	0.28	&	63.85	&	0.75	$\pm$	0.18	&	1$^{\rm b}$	&	--	\\
E2	&	5.23	$\pm$	0.78	&	10.21	$\pm$	0.21	&	67.04	&	1.37	$\pm$	0.29	&	1$^{\rm b}$	&	--	\\
E1	&	3.92	$\pm$	0.59	&	12.94	$\pm$	0.46	&	64.45	&	0.77	$\pm$	0.18	&	1$^{\rm b}$	&	--	\\
\noalign{\smallskip}
\multicolumn{7}{c}{ \it \hrulefill 2004.80 \hrulefill } \\
\noalign{\smallskip}
0	&	1590.66	$\pm$	238.60	&	0.00	&	--	&	0.28	$\pm$	0.11	&	0.38	&	66.10	\\
N	&	507.49	$\pm$	76.12	&	0.33	$\pm$	0.15	&	61.95	&	0.14	$\pm$	0.10	&	1$^{\rm b}$	&	--	\\
M	&	294.28	$\pm$	44.14	&	0.56	$\pm$	0.15	&	70.23	&	0.22	$\pm$	0.11	&	1$^{\rm b}$	&	--	\\
L	&	34.00	$\pm$	5.10	&	1.06	$\pm$	0.20	&	65.05	&	0.30	$\pm$	0.12	&	1$^{\rm b}$	&	--	\\
K	&	15.62	$\pm$	2.34	&	1.75	$\pm$	0.20	&	70.84	&	0.38	$\pm$	0.13	&	1$^{\rm b}$	&	--	\\
J	&	13.76	$\pm$	2.06	&	2.55	$\pm$	0.28	&	66.12	&	0.33	$\pm$	0.12	&	1$^{\rm b}$	&	--	\\
I	&	20.41	$\pm$	3.06	&	3.62	$\pm$	0.22	&	64.10	&	0.62	$\pm$	0.16	&	1$^{\rm b}$	&	--	\\
H4	&	41.27	$\pm$	6.19	&	4.93	$\pm$	0.24	&	63.72	&	0.64	$\pm$	0.16	&	1$^{\rm b}$	&	--	\\
H3	&	88.46	$\pm$	13.27	&	5.25	$\pm$	0.21	&	62.60	&	0.87	$\pm$	0.20	&	1$^{\rm b}$	&	--	\\
H2	&	97.12	$\pm$	14.57	&	5.59	$\pm$	0.18	&	59.49	&	1.10	$\pm$	0.24	&	1$^{\rm b}$	&	--	\\
H1	&	13.27	$\pm$	1.99	&	6.47	$\pm$	0.24	&	61.41	&	1.82	$\pm$	0.38	&	1$^{\rm b}$	&	--	\\
G	&	21.67	$\pm$	3.25	&	7.80	$\pm$	0.07	&	61.58	&	0.37	$\pm$	0.12	&	1$^{\rm b}$	&	--	\\
E3	&	4.32	$\pm$	0.65	&	9.85	$\pm$	0.28	&	66.41	&	0.97	$\pm$	0.22	&	1$^{\rm b}$	&	--	\\
E1	&	6.74	$\pm$	1.01	&	13.19	$\pm$	0.46	&	64.15	&	0.77	$\pm$	0.18	&	1$^{\rm b}$	&	--	\\
\noalign{\smallskip}
\multicolumn{7}{c}{ \it \hrulefill 2005.02 \hrulefill } \\
\noalign{\smallskip}
0	&	1631.40	$\pm$	244.71	&	0.00	&	--	&	0.24	$\pm$	0.11	&	0.21	&	61.51	\\
N	&	542.58	$\pm$	81.39	&	0.31	$\pm$	0.15	&	70.03	&	0.06	$\pm$	0.10	&	1$^{\rm b}$	&	--	\\
M	&	386.90	$\pm$	58.03	&	0.67	$\pm$	0.15	&	63.61	&	0.15	$\pm$	0.10	&	1$^{\rm b}$	&	--	\\
X11	&	187.02	$\pm$	28.05	&	0.86	$\pm$	0.30	&	70.85	&	0.20	$\pm$	0.11	&	1$^{\rm b}$	&	--	\\
L	&	17.02	$\pm$	2.55	&	1.47	$\pm$	0.20	&	67.37	&	-$^{\rm c}$			&	1$^{\rm b}$	&	--	\\
K	&	12.09	$\pm$	1.81	&	1.99	$\pm$	0.20	&	68.81	&	0.39	$\pm$	0.13	&	1$^{\rm b}$	&	--	\\
J	&	16.23	$\pm$	2.43	&	3.21	$\pm$	0.28	&	64.59	&	0.99	$\pm$	0.22	&	1$^{\rm b}$	&	--	\\
I	&	12.46	$\pm$	1.87	&	4.28	$\pm$	0.22	&	64.94	&	0.54	$\pm$	0.15	&	1$^{\rm b}$	&	--	\\
H4	&	101.42	$\pm$	15.21	&	5.42	$\pm$	0.24	&	62.82	&	0.47	$\pm$	0.14	&	1$^{\rm b}$	&	--	\\
H3	&	58.52	$\pm$	8.78	&	5.81	$\pm$	0.21	&	59.93	&	0.39	$\pm$	0.13	&	1$^{\rm b}$	&	--	\\
H2	&	46.72	$\pm$	7.01	&	6.18	$\pm$	0.18	&	59.53	&	0.79	$\pm$	0.19	&	1$^{\rm b}$	&	--	\\
H1	&	4.59	$\pm$	0.69	&	7.16	$\pm$	0.24	&	61.98	&	-$^{\rm c}$			&	1$^{\rm b}$	&	--	\\
G	&	15.88	$\pm$	2.38	&	8.07	$\pm$	0.07	&	61.54	&	0.87	$\pm$	0.20	&	1$^{\rm b}$	&	--	\\
E3	&	4.09	$\pm$	0.61	&	10.02	$\pm$	0.28	&	66.09	&	1.51	$\pm$	0.32	&	1$^{\rm b}$	&	--	\\
E1	&	5.04	$\pm$	0.76	&	13.35	$\pm$	0.46	&	65.18	&	1.28	$\pm$	0.27	&	1$^{\rm b}$	&	--	\\
\noalign{\smallskip}
\multicolumn{7}{c}{ \it \hrulefill 2005.73 \hrulefill } \\
\noalign{\smallskip}
0	&	1197.32	$\pm$	179.60	&	0.00	&	--	&	0.28	$\pm$	0.11	&	0.30	&	69.16	\\
X14	&	480.43	$\pm$	72.06	&	0.35	$\pm$	0.30	&	70.12	&	0.12	$\pm$	0.10	&	1$^{\rm b}$	&	--	\\
X13	&	224.64	$\pm$	33.70	&	0.67	$\pm$	0.30	&	68.78	&	0.13	$\pm$	0.10	&	1$^{\rm b}$	&	--	\\
N	&	71.31	$\pm$	10.70	&	1.07	$\pm$	0.15	&	63.01	&	0.23	$\pm$	0.11	&	1$^{\rm b}$	&	--	\\
M	&	219.35	$\pm$	32.90	&	1.86	$\pm$	0.15	&	69.54	&	0.34	$\pm$	0.12	&	0.67	&	41.38	\\
L	&	13.95	$\pm$	2.09	&	2.41	$\pm$	0.20	&	65.06	&	0.29	$\pm$	0.12	&	1$^{\rm b}$	&	--	\\
K	&	4.81	$\pm$	0.72	&	3.09	$\pm$	0.20	&	66.29	&	-$^{\rm c}$			&	1$^{\rm b}$	&	--	\\
J	&	12.38	$\pm$	1.86	&	4.32	$\pm$	0.28	&	64.25	&	0.78	$\pm$	0.18	&	1$^{\rm b}$	&	--	\\
I	&	12.70	$\pm$	1.91	&	5.33	$\pm$	0.22	&	64.91	&	0.85	$\pm$	0.20	&	1$^{\rm b}$	&	--	\\
H4	&	32.74	$\pm$	4.91	&	6.20	$\pm$	0.24	&	61.84	&	0.46	$\pm$	0.14	&	1$^{\rm b}$	&	--	\\
H3	&	38.49	$\pm$	5.77	&	6.60	$\pm$	0.21	&	60.79	&	0.53	$\pm$	0.15	&	1$^{\rm b}$	&	--	\\
H2	&	35.44	$\pm$	5.32	&	7.21	$\pm$	0.18	&	58.47	&	0.84	$\pm$	0.19	&	1$^{\rm b}$	&	--	\\
H1	&	13.71	$\pm$	2.06	&	8.20	$\pm$	0.24	&	63.77	&	0.88	$\pm$	0.20	&	1$^{\rm b}$	&	--	\\
G	&	5.03	$\pm$	0.75	&	9.56	$\pm$	0.07	&	59.78	&	0.52	$\pm$	0.14	&	1$^{\rm b}$	&	--	\\
E3	&	1.97	$\pm$	0.30	&	11.42	$\pm$	0.28	&	66.32	&	-$^{\rm c}$			&	1$^{\rm b}$	&	--	\\
E1	&	4.99	$\pm$	0.75	&	15.30	$\pm$	0.46	&	64.80	&	2.10	$\pm$	0.43	&	1$^{\rm b}$	&	--	\\
\vspace{2.0ex} \\
\enddata
\tablenotetext{a}{The PA is measured from north through east}
\tablenotetext{b}{Axis ratio fixed at 1}
\tablenotetext{c}{Undetermined by the fit}
\end{deluxetable}

\clearpage
\begin{table}
      \caption[]{Kinematics
         \label{tab:3c111_kinematics}}
\[
\centering
\resizebox{0.8\columnwidth}{!}{%
            \begin{tabular}{@{}cccccc@{}}
\hline
\hline
\noalign{\smallskip}
Component  & \# of epochs & $\mu$ & $\beta_{\rm app}$ & Peak Flux  & Ejection \\
           &              & [mas\,yr$^{-1}$] &         & [Jy] & Epoch \\
            \noalign{\smallskip}
\hline
\noalign{\smallskip}
B & 4 & $1.74 \pm 0.04$ & $5.7 \pm 0.1$ & -- & $1994.07 \pm 0.06$\\
\noalign{\smallskip}
C & 4 & $1.4 \pm 0.1$ & $4.6 \pm 0.3$ & -- & $1994.60 \pm 0.19$ \\
\noalign{\smallskip}
E & 14 & $1.00 \pm 0.02$ & $3.26 \pm 0.07$ & 1.46 & $1996.10 \pm 0.03$\\
\noalign{\smallskip} 
E\,1 & 12 & $1.69 \pm 0.04$ & $5.5 \pm 0.1$ & 0.30 & --\\
\noalign{\smallskip}
E\,2 & 7 & $1.29 \pm 0.06$ & $4.2 \pm 0.2$ & 0.13 & -- \\
\noalign{\smallskip}
E\,3 & 12 & $1.22 \pm 0.05$ & $4.0 \pm 0.2$ & 0.47 & -- \\
\noalign{\smallskip}
E\,4 & 7 & $0.86 \pm 0.05$ & $2.8 \pm 0.2$ & 0.12 & -- \\
\noalign{\smallskip}
F & 4 & $0.64 \pm 0.07$ & $2.1 \pm 0.2$ & 1.98 & $1996.13 \pm 0.16$\\
\noalign{\smallskip} 
G & 9 & $1.83 \pm 0.02$ & $5.97 \pm 0.07$ & 0.24 & $2000.54 \pm 0.03$\\
\noalign{\smallskip}
H& 4 & $1.73 \pm 0.02$& $5.64 \pm 0.07$ & 0.43 & $2001.43 \pm 0.02$ \\
\noalign{\smallskip}
H1& 4 & $1.9 \pm 0.2$ & $6.2  \pm 0.7$ & 0.03 & -- \\
\noalign{\smallskip}
H2& 4 & $1.71 \pm 0.11$& $5.6 \pm 0.4$ & 0.10 & -- \\
\noalign{\smallskip} 
H3& 4 & $1.5 \pm 0.1$& $4.9 \pm 0.3$ & 0.11 & -- \\
\noalign{\smallskip}
H4& 3 & $1.3 \pm 0.2$ & $4.2 \pm 0.7$ & 0.10 & -- \\
\noalign{\smallskip}
I & 6 & $1.3 \pm 0.1$ & $4.2 \pm 0.4$ & 0.08 & $2001.77 \pm 0.21$ \\
\noalign{\smallskip}
J & 6 & $1.2 \pm 0.1$ & $3.9 \pm 0.3$ & 0.10 & $2002.30 \pm 0.24$ \\
\noalign{\smallskip}
K & 5 & $1.1 \pm 0.1$ & $3.6 \pm 0.7$ & 0.18 & $2003.08 \pm 0.22$ \\
\noalign{\smallskip}
L & 4 & $1.1 \pm 0.2$ & $3.6 \pm 0.7$ & 0.11 &  $2003.74 \pm 0.22$ \\
\noalign{\smallskip}
M & 3 & $1.5 \pm 0.2$ & $4.9 \pm 0.7$ & 0.39 & $2004.50 \pm 0.12$ \\ 
\noalign{\smallskip}
N & 3 & $0.9 \pm 0.2$ & $2.9 \pm 0.7$ & 0.54 &  $2004.57 \pm 0.19$ \\
\noalign{\smallskip}
\hline
\end{tabular}
}
\]
\end{table}


\clearpage
\begin{thebibliography}{}
%
\bibitem[Agudo et al.(2001)]{Agu01} Agudo, I., G\'omez, J.,
Mart{\'{\i}}, J.~M., et al.\
2001, \apj, 549, L183
%
\bibitem[Alef et al.(1998)]{Ale98} Alef, W., Preuss, E., Kellermann,
K.~I., Gabuzda, D.\ 1998, in Radio Emission from Galactic and Extragalactic Compact Sources, ASP Conf. Ser. 144, 129
%
\bibitem[Aller, Aller, \& Hughes(2003)]{All03} Aller, M.~F., Aller,
H.~D., Hughes, P.~A.\ 2003, in
Radio Astronomy at the Fringe, Zensus, J.~A., Cohen, M.~H., Ros, E. (eds.),
ASP Conference Ser. 300, 159
%
\bibitem[Aloy et al.(2003)]{Alo03} Aloy, M.-{\'A}.,
Mart{\'{\i}}, J.-M., G{\'o}mez, J.-L., Agudo, I., M{\"u}ller, E., \&
Ib{\'a}{\~n}ez, J.-M.\ 2003, \apjl, 585, L109
%
\bibitem[Blandford \& Konigl(1979)]{Bla79} Blandford, R.~D., 
\& Konigl, A.\ 1979, \apj, 232, 34
%
\bibitem[Cara \& Lister(2007)]{Car07} Cara, M., \& Lister, 
M.~L.\ 2007, ArXiv Astrophysics e-prints, arXiv:astro-ph/0702449
%
\bibitem[Cohen et al.(1971)]{Coh71} Cohen, M.~H., Cannon, W., 
Purcell, G.~H., Shaffer, D.~B., Broderick, J.~J., Kellermann, K.~I., \& 
Jauncey, D.~L.\ 1971, \apj, 170, 207
%
\bibitem[Cohen et al.(1977)]{Coh77} Cohen, M.~H., et al.\ 
1977, \nat, 268, 405 
%
\bibitem[Cohen et al.(2007)]{Coh07} Cohen, M.~H., Lister, 
M.~L., Homan, D.~C., Kadler, M., Kellermann, K.~I., Kovalev, Y.~Y., \& 
Vermeulen, R.~C.\ 2007, \apj, 658, 232
%
\bibitem[Goetz et al.(1987)]{Goe87}
Goetz, M.~M.~A., Preuss, E., Alef, W., \& Kellermann, K.~I.\ 1987, \aap,
176, 171
%
\bibitem[G\'omez et al.(1997)]{Gom97} G\'omez, J.~L., Mart\'{\i}, 
J.~M.~A., Marscher, A.~P., Ibanez, J.~M.~A., \& Alberdi, A.\ 1997, \apjl, 
482, L33
%
\bibitem[G\'omez et al.(2001)]{Gom01} G\'omez, J., Marscher, A.~P.,
Alberdi, A., Jorstad, S.~G., Agudo, I.\ 2001, \apjl, 561, L161
%
\bibitem[G{\'o}mez(2005)]{Gom05} G{\'o}mez, J.\ 2005, Future 
Directions in High Resolution Astronomy, 340, 13 
%
\bibitem[Hartman et al.(1999)]{Har99} Hartman, R.~C., et al.\ 
1999, \apjs, 123, 79
%
\bibitem[Hartman \& Kadler(2007)]{Har07} Hartman, R.~C. \& Kadler, M.\ 2007, \apj, submitted
%
\bibitem[Homan et al.(2006)]{Hom06} Homan, D.~C., et al.\ 
2006, \apjl, 642, L115
%
\bibitem[Homan \& Lister(2006)]{Hom06b} Homan, D.~C., \& 
Lister, M.~L.\ 2006, \aj, 131, 1262
%
\bibitem[Homan et al.(2003)]{Hom03} Homan, D.~C., Lister, 
M.~L., Kellermann, K.~I., Cohen, M.~H., Ros, E., Zensus, J.~A., Kadler, M., 
\& Vermeulen, R.~C.\ 2003, \apjl, 589, L9 
%
\bibitem[Homan et al.(2002)]{Hom02} Homan, D.~C., Ojha, R., 
Wardle, J.~F.~C., Roberts, D.~H., Aller, M.~F., Aller, H.~D., \& Hughes, 
P.~A.\ 2002, \apj, 568, 99
%
\bibitem[Hughes et al.(1985)]{Hug85} Hughes, P.~A., Aller, 
H.~D., \& Aller, M.~F.\ 1985, \apj, 298, 301
%
\bibitem[Jorstad et al.(2001)]{Jor01} Jorstad, S.~G., 
Marscher, A.~P., Mattox, J.~R., Wehrle, A.~E., Bloom, S.~D., \& Yurchenko, 
A.~V.\ 2001, \apjs, 134, 181 
%
\bibitem[Jorstad et al.(2005)]{Jor05} Jorstad, S.\,G., Marscher, A.\,P.,
Lister, M.\,L., et al.\ 2005, \aj, 130, 1418
%
\bibitem[Kadler(2005)]{Kad05} Kadler, M.\ 2005, Ph.\,D. Thesis, Rheinische Friedrich-Wilhelms-Universit\"at Bonn, Bonn, Germany
%
\bibitem[Kadler et al.(2004)]{Kad04} Kadler, M., Ros, E., Lobanov, A.\,P., Falcke, H., Zensus, J.\,A.\ 2004, \aap, 426, 481
%
\bibitem[Kellermann et al.(1998)]{Kel98} Kellermann, K.~I., Vermeulen, R.~C.,
Zensus, J.~A., \& Cohen, M.~H.\ 1998, \aj, 115, 1295
%
\bibitem[Kellermann \& Moran(2001)]{Kel01} Kellermann, K.~I., 
\& Moran, J.~M.\ 2001, \araa, 39, 457
%
\bibitem[Kellermann et al.(2004)]{Kel04} Kellermann, K.~I., Lister,
M.~L., Homan, D.~C., et al.\ 2004, \apj, 609, 539
%
%
\bibitem[Kovalev et al.(2005)]{Kov05} Kovalev, Y.~Y., Kellermann, K.~I., Lister, M.~L.,
et al.\ 2005, \aj, 130, 2473
%
\bibitem[L\"ahteenm\"aki \& Valtaoja(1999)]{Lae99} L\"ahteenm\"aki, A.
\& Valtaoja, E.\ 1999, \apj, 521, 493
%
\bibitem[Leismann et al.(2005)]{Lei05} Leismann, T., 
Ant{\'o}n, L., Aloy, M.~A., M{\"u}ller, E., Mart{\'{\i}}, J.~M., Miralles, 
J.~A., \& Ib{\'a}{\~n}ez, J.~M.\ 2005, \aap, 436, 503 
%
\bibitem[Lewis et al.(2005)]{Lew05} Lewis, K.\,T., Eracleous, M.,
Gliozzi, M., et al.\ 2005, \apj, 622, 816
%
\bibitem[Linfield \& Perley(1984)]{Lin84} Linfield, R. \& Perley, R.\ 1984, \apj, 279, 60
%
\bibitem[Lister \& Homan(2005)]{Lis05} Lister, M.~L., \&
Homan, D.~C.\ 2005, \aj, 130, 1389
%
\bibitem[Lott et al.(2007)]{Lot07} Lott, B., Carson, J.,
Ciprini, S., Dermer, C.~D., Giommi, P., Madejski, G., Lonjou, V., \&
Reimer, A.\ 2007, American Institute of Physics Conference Series, 921, 347 
%
\bibitem[Marscher \& Gear(1985)]{Mar85} Marscher, A.~P., \& 
Gear, W.~K.\ 1985, \apj, 298, 114 
%
\bibitem[Marscher et al.(2002)]{Mar02} Marscher, A.~P., 
Jorstad, S.~G., G{\'o}mez, J.-L., Aller, M.~F., Ter{\"a}sranta, H., Lister, 
M.~L., \& Stirling, A.~M.\ 2002, \nat, 417, 625 
%
\bibitem[Marscher (2006)]{Mar06} Marscher, A.~P.\ 2006, AN 327, 217
%
\bibitem[Mukherjee et al.(2002)]{Muk02} Mukherjee, R., 
Halpern, J., Mirabal, N., \& Gotthelf, E.~V.\ 2002, \apj, 574, 693 
%
\bibitem[Mimica et al.(2004)]{Mim04} Mimica, P., Aloy, M.~A., 
M{\"u}ller, E., \& Brinkmann, W.\ 2004, \aap, 418, 947 
%
\bibitem[Mimica et al.(2007)]{Mim07} Mimica, P., Aloy, M.~A., 
M\"uller, E.\ 2007, \aap, 466, 93 
%
\bibitem[Mizuno et al.(2007)]{Miz07} Mizuno, Y., Hardee, P., 
\& Nishikawa, K.-I.\ 2007, \apj, 662, 835
%
%
\bibitem[Perucho et al.(2006)]{Per06} Perucho, M., Lobanov, A.P.,
Mart\'{\i}, J.M., Hardee, P.E.\ A\&A 456, 493, 2006
%
\bibitem[Perucho et al.(2007)]{Per07} Perucho, M., Hanasz, M., Mart\'{\i},
J.M., Miralles, J.A.\ 2007, Phys. Rev. E, 75, 056312
%
\bibitem[Perucho \& Lobanov(2007)]{Per07a} Perucho, M., \& Lobanov,
A.~P.\ 2007, \aap, 469, L23
%
\bibitem[Perucho \& Mart\'{\i}(2007)]{Per07b} Perucho, M., \&
Mart\'{\i}, J.M.\ 2007, MNRAS, 382, 526
%
\bibitem[Piner et al.(2007)]{Pin07} Piner, B.~G., Mahmud, M., 
Fey, A.~L., \& Gospodinova, K.\ 2007, \aj, 133, 2357
%
\bibitem[Preuss, Alef, \& Kellermann(1988)]{Pre88} Preuss, E., Alef, W.,
Kellermann, K.~I.\ 1988, in
IAU Symp.~129: The Impact of VLBI on Astrophysics and Geophysics, 105
%
\bibitem[Roca-Sogorb et al.(2008)]{Roc08} Roca-Sogorb, M., Perucho, M.,
Gómez, J.L., Martí, J.M., Antón, L., Aloy, M.A., Agudo, I.\ 2008, in
proceedings of Extragalactic Jets: Theory and Observation from Radio to
Gamma-Ray, eds: T.A.~Rector, D.S.~de~Young, ASP Conference Series, in
press
%
%
\bibitem[Scheck et al.(2002)]{Sch02} Scheck, L., Aloy, M.~A., 
Mart{\'{\i}}, J.~M., G{\'o}mez, J.~L., M\"uller, E.\ 2002, \mnras, 
331, 615 
%
\bibitem[Sguera et al.(2005)]{Sgu05} Sguera, V., Bassani, L., 
Malizia, A., Dean, A.~J., Landi, R., \& Stephen, J.~B.\ 2005, \aap, 430, 
107 
%
\bibitem[Shepherd(1997)]{She97} Shepherd, M.~C.\ 1997, 
Astronomical Data Analysis Software and Systems VI, 125, 77 
%
\bibitem[Sreekumar et al.(1999)]{Sre99} Sreekumar, P., 
Bertsch, D.~L., Hartman, R.~C., Nolan, P.~L., \& Thompson, D.~J.\ 1999, 
Astroparticle Physics, 11, 221
%
\bibitem[Ter\"asranta et al.(2004)]{Ter04} Ter\"asranta, H., Achren, J.,
Hanski, M., et al.\ 2004, \aap, 427, 769
%
\bibitem[Tingay, Preston, \& Jauncey(2001)]{Tin01} Tingay, S.~J., Preston, R.~A., \& Jauncey, D.~L.\ 2001, \aj, 122, 1697
%
\bibitem[Vermeulen \& Cohen(1994)]{Ver94} Vermeulen, R.~C., 
\& Cohen, M.~H.\ 1994, \apj, 430, 467
%
\bibitem[Whitney et al.(1971)]{Whi71} Whitney, A.~R., et al.\ 
1971, Science, 173, 225
%
\bibitem[Zavala \& Taylor(2002)]{Zav02} Zavala, R.~T., \&
Taylor, G.~B.\ 2002, \apjl, 566, L9
%
\bibitem[Zensus(1997)]{Zen97} Zensus, J.~A.\ 1997, \araa, 35, 
607
%
\bibitem[Zensus et al.(2002)]{Zen02} Zensus, J.~A., Ros, E.,
Kellermann, K.~I., et al.\ 2002, \aj, 124, 662
%
\end{thebibliography}
\end{document}